\RequirePackage{lineno}
\documentclass[aps, prstab, preprint, groupedaddress, longbibliography]{revtex4-1}

\usepackage[utf8]{inputenc}
\usepackage{lineno}
\usepackage{hyperref}
\usepackage{upgreek}
\usepackage[colorinlistoftodos]{todonotes}

\begin{document}
\begin{center}
  \textbf{\large Opportunities in Machine Learning for Particle Accelerators}\\[3ex]
  \rule[-3ex]{40ex}{1pt}\\[0.5ex]Editors\\\rule[2.5ex]{40ex}{1pt}\\
  A. Edelen and C. Mayes\\
  \emph{SLAC National Accelerator Laboratory, Menlo Park, CA 94025, USA}\\[2ex]
  D. Bowring\\[1ex]
  \emph{Fermi National Accelerator Laboratory, Batavia, IL 60510, USA}\\[1ex]
  \rule[-3ex]{40ex}{1pt}\\[0.5ex]Contributors\\\rule[2.5ex]{40ex}{1pt}\\
  D. Ratner\\
  \emph{SLAC National Accelerator Laboratory, Menlo Park, CA 94025, USA}\\[2ex]
  A. Adelmann, R. Ischebeck, and J. Snuverink\\
  \emph{Paul Scherrer Institut, Villigen, Switzerland}\\[2ex]
  I. Agapov and R. Kammering\\
  \emph{Deutsches Elektronen-Synchrotron, Hamburg, Germany}\\[2ex]
  J. Edelen\\
  \emph{Radiasoft, LLC, Boulder, Colorado 80301, USA}\\[2ex]
  I. Bazarov\\
  \emph{Cornell University, Ithaca, NY 14853, USA}\\[2ex]
  G. Valentino\\
  \emph{University of Malta, Msida, Malta}\\[2ex]
  J. Wenninger\\
  \emph{CERN, Geneva, Switzerland}
\end{center}

\newpage\begin{abstract}
``Machine learning'' (ML) is a subfield of artificial intelligence.  The term applies broadly to a collection of computational algorithms and techniques that train systems from raw data rather than \emph{a priori} models. ML techniques are now technologically mature enough to be applied to particle accelerators, and we expect that ML will become an increasingly valuable tool to meet new demands for beam energy, brightness, and stability. The intent of this white paper is to provide a high-level introduction to problems in accelerator science and operation where incorporating ML-based approaches may provide significant benefit. We review ML techniques currently being investigated at particle accelerator facilities, and we place specific emphasis on active research efforts and promising exploratory results. We also identify new applications and discuss their feasibility, along with the required data and infrastructure strategies.  We conclude with a set of guidelines and recommendations for laboratory managers and administrators, emphasizing the logistical and technological requirements for successfully adopting this technology. This white paper also serves as a summary of the discussion from a recent workshop held at SLAC on ML for particle accelerators ~\cite{mlworkshop2018}. 
\end{abstract}
\maketitle

\section{Introduction}
Machine learning (ML) is a subfield of artificial intelligence centered on algorithms that can learn to complete tasks using data.
ML is currently experiencing a renaissance thanks to recent computational, theoretical, and practical advances in the field. Recently, various ML approaches have been successfully demonstrated in real-world tasks for computer vision~\cite{DBLP:journals/nature/LeCunBH15}, anomaly 
detection~\cite{Awoyemi2017}, and language translation ~\cite{Hirschberg:2015}. 
Those wishing to use ML are able to access an increasingly wide variety of high performance computing resources. For example, computing clusters at universities and other research institutions are now widely accessible, along with a variety of web-based services.
The greater availability of large data sets and the variety of new open source high-level software libraries has also helped to improve the reach of existing ML techniques and to facilitate rapid prototyping of new algorithms.
These technical advances go hand-in-hand with new real-world applications that motivate ML research and funding in the public and private sectors, as well as provide feedback that informs further algorithmic development.

Particle accelerators in particular are some of the largest, most data-intensive, and most complex scientific systems  in existence. The interrelations between machine subsystems are complicated and often nonlinear, the system dynamics involve large parameter spaces that evolve over multiple relevant time scales, and accelerator systems can be difficult to model \emph{a priori}. Relevant problems for accelerators include, for example, analysis of large quantities of archived data, accurate and fast modeling of accelerator systems, detection of aberrant machine behavior, optimization of accelerator design, and active tuning and control. At present, ML-based approaches are technologically mature enough to be brought to bear on a wide variety of problems within these domains.

We expect that ML will become an increasingly valuable tool to meet new demands for beam energy, brightness, reliability, and stability. In this white paper we review ML techniques currently being investigated at particle accelerator facilities, including several promising experimental demonstrations. We also identify new applications and discuss their feasibility, along with the required data and infrastructure strategies.

Much of the content here has been guided by presentations and discussion arising from the inaugural International Committee for Future Accelerators (ICFA) Workshop on Machine Learning for Particle Accelerators, held from February 27 through March 2, 2018 at SLAC National Accelerator Laboratory~\cite{mlworkshop2018}. A similar workshop was held at Daresbury Lab on January 30th and 31st, 2018~\cite{daresburyworkshop2018}. For some additional introductory material on the topic of ML for particle accelerators, see ~\cite{Edelen16}.

\clearpage
\section{General applications of machine learning}\label{sec:otheruses}
ML has been successfully applied to a variety of other scientific/engineering
problems. Here we review some applications beyond accelerator science for historical context.

The modern wave of ML algorithms (e.g. deep neural networks), first began to see success in computer vision applications. For a high-level overview of some recent successes in classic computer vision domains, e.g. object recognition in natural scenes, see Refs.~\cite{DBLP:journals/nature/LeCunBH15,SCHMIDHUBER201585}.  
In astronomy, neural networks have been used extensively for automated sorting and analysis of sky survey images and rejection of artifacts~\cite{doi:10.1142/S0218271810017160, way2017, Graff2013, doi:10.1093/mnras/stu642, PhysRevD.88.062003}. 
In particle detectors, ML methods are used to identify and  classify particle tracks ~\cite{pbhat}.
More recently, this has included deep neural networks~\cite{AAurisano:2016, Baldi2014SearchingFE, doi:10.1146/annurev.nucl.012809.104427, WHITESON20091203}, and a review of these developments can be found in ~\cite{Radovic}. Notably, a 2014 ``Kaggle'' code competition also drew numerous solutions from the public to identify ATLAS detector signals related to the Higgs boson. Contestants used neural networks, decision trees, and ensemble methods ~\cite{Adam-Bourdarios:2015pye}.

In the medical community, ML has been used for automated diagnosis of medical conditions. ML algorithms have recently shown improvements over traditional screening techniques in identifying/classifying skin and cervical cancers\cite{DBLP:journals/nature/EstevaKNKSBT17, Zhang04}. Other efforts showed that a ML algorithm was able to beat cardiologists at detecting heart arrhythmia viewed on an electrocardiogram~\cite{2017arXiv170701836R}

As an industrial example, robotic arms have been controlled using ML algorithms to perform sorting and packaging operations. These algorithms have been used reliably in operation for years ~\cite{ROB-021}. Additionally, ML has been in use by the U.S. post office for the interpretation of hand-written addresses since the 1990s with an accuracy of above 98\%, saving 100 million dollars during the first year of deployment~\cite{Srihari1997}

With regard to anomaly detection and machine protection, the Joint European Torus has used  ML for instability detection~\cite{Murari2012,Murari2012IdentifyingJI} and fault prediction~\cite{Murari2012IdentifyingJI,MURARI20132} in its fusion research facility. 
ML was also recently used to optimize tokamak settings for plasma confinement, resulting in over 50 percent improvement in key performance metrics such as energy loss rate and total plasma energy ~\cite{optometrist-algorithm}.

\clearpage
\section{Early history of usage for particle accelerators}
ML techniques have been applied to particle accelerators since the late 1980s. 
Much early discussion during the late 1980s and early 1990s focused on applying rule-based systems to accelerator control and tuning \cite{Higo86, Weygand87, Skarek96, Schultz90, Fiesler99}.
In the early 1990s, scientists at Los Alamos National Lab had some experimental success with neural-network-based ion source control \cite{Howell90, Mead92, Mead94}. 
Other early studies at the University of New Mexico focused on orbit/trajectory control \cite{Bozoki94,Nguyen91,Hitaka04,Kim00,Schirmer06}, fault detection and management \cite{Jennings96,Perriollat96}, and root-cause analysis of  
errors (e.g. distinguishing between bending magnet misalignment and field defects ~\cite{Kijima92}). General AI/ML platform for beamline tuning were also planned \cite{Stern97,Klein97,Klein97-2,Klein99}.  None of these systems were eventually used routinely as part of an accelerator's main control system.

The lack of clear success in bringing ML to regular use in accelerator systems was partly due to limitations in the then-available hardware, algorithms, and software packages, as well as the limited accessibility of good data sets and simulation tools. Similar situations were encountered in other scientific fields where ML approaches were tried before they had reached sufficient technological maturity relative to the challenges of the particular application.

\clearpage
\section{Areas of applicability for particle accelerators}
ML-based solutions to challenges encountered in particle accelerators are now under development and yielding promising results. As ML becomes a routine tool in this field, other applications are likely to emerge. Here we review several use cases.

\subsection{Anomaly detection and machine protection}
The operation of today’s particle accelerators is increasingly complex and data-intensive,  and it typically involves a large number of process variables, nonlinear behavior, and many interacting subsystems. The identification of conditions that could negatively impact machine performance requires monitoring of many interacting subsystems, and conventional analysis methods are often insufficient.  Using machine data to identify poor conditions and the root cause of errors encountered during operation is therefore a growing challenge. In addition, improving the reliability of machine protection systems can help pave the way to producing beams of higher intensity/stored energy. 

Anomaly and breakout detection algorithms may provide added capabilities beyond those of existing machine protection systems by detecting subtle behaviors of key variables prior to negative events. This approach has been used for detection of quench precursors in both superconducting magnets~\cite{Wielgosz:2016xhl} and superconducting RF cavities~\cite{NawazDESY2016}.

In the context of system control, anomaly detection algorithms can also be used to identify and throw away bad signals. The LHC has used such algorithms to identify bad readings from beam position monitors and remove them prior to a standard correction procedure~\cite{Fol:2309558}, as well as to assist in automated collimator alignment~\cite{Valentino:2017hlm}. 

\subsection{System Modeling}
Another challenge in accelerator science is the need to obtain machine models that (a) are sufficiently accurate and (b) can execute quickly enough to be useful during operations, possibly with input from machine instrumentation (i.e. \textbf{online modeling}), and in large-scale offline optimization.

Many simulation tools are too slow in practice to be used directly in control systems or to provide guidance to operators during machine operation.
In recent years, some hardware-accelerated modeling codes have been developed that only require modest computing resources, for example, GPU-accelerated \texttt{elegant}~\cite{King:2017vfo} and \texttt{HPSim}, a GPU-accelerated code for online modeling of ion linacs~\cite{Pang:2014cda,Pang:2015iul}.
However, most online models at present rely on simplified representations of the relevant physics: accuracy is traded for speed. Furthermore, ensuring these models accurately represent the behavior of a particular machine still requires substantial work in most cases and is not always possible due to practical constraints (e.g. lack of diagnostics, lack of personnel, lack of time for systematic machine studies).

System modeling with ML enables one to learn representations that combine information from physics-based simulations with measured data, which can be useful when measured data only covers a limited portion of the system behavior that needs to be included for prediction or control. This has been demonstrated for the injector and low-energy beamline at Fermilab's FAST facility ~\cite{Edel18_2}. Furthermore, the execution speed of such ML models can be faster than conventional accelerator simulations, enabling fast-executing representations of slowly-executing, high-fidelity simulations to be created.  This technique is known as \textbf{surrogate modeling}. For example, in one early study a neural network surrogate model of the RF gun and low-energy beamline at FAST sped up the simulation time from around 20 minutes to under a millisecond~\cite{Edelen:2017ewy}.
The resulting simulation tools can be useful for providing real-time results in a control room environment or for planning experiments and doing offline controller design. For some examples, see ~\cite{Adelmann:2015wva,Edelen:2017ewy,auraleeFEL2017}. 

Neural networks in particular are also well-suited to tasks which deal with image data, and this  enables image-based diagnostics to be used directly in accelerator models, both as outputs and inputs. For example, to take into account variability in a drive laser for a photocathode gun, a virtual cathode image of the laser spot has been used as one input to a model to predict downstream beam parameters at FAST ~\cite{Edelen:2017ewy}.

In terms of offline optimization and design, current approaches take advantage of heavy parallelization when a high-fidelity physics model is required (for example see ~\cite{pelegant,opal_parallel}). However, gaining access to the necessary computing resources for this is not always possible, and in practice simulation speed often limits the extent of parameter space exploration during the optimization and design process. In this case, ML models that are sufficiently well-validated could be used instead to interpolate across optimum fronts of accelerator performance from a set of sparsely-sampled training examples. This ultimately could lead to an increase in the throughput of optimization for challenging accelerator design problems (such as dynamic aperture maximization for diffraction-limited storage ring light source ~\cite{Huang:2015wka}) or allow faster exploration of various trade-offs between competing parameters. 

\subsection{Virtual Instrumentation / Virtual Diagnostics}
Accelerators are outfitted with a variety of instruments to provide measurements of both the beam itself and other key variables that impact the beam (e.g. RF signals). These instruments are often referred to as "diagnostics," in the sense that they provide some indication of the performance and operating state of the accelerator. A \textbf{virtual instrument (also referred to as a "virtual diagnostic")} provides an estimate of what an instrument would read when such a reading is unavailable, thus allowing non-invasive  "measurements" in locations where it may be impossible to put an instrument, where only destructive instruments are available and thus cannot be used during normal operations, where the update rate of an existing instrument is lower than needed for experimental analysis or for control, or where the instrument is not sensitive across the entire operating range. 

For example, a virtual instrument for the beam phase space based on high-fidelity, GPU-accelerated physics simulations has been demonstrated at LANSCE. The simulation reads inputs from the machine and within seconds provides predictions about the resulting beam, including live plots of the phase space ~\cite{Pang:2015iul}. In a slightly different approach demonstrated at FACET, free parameters in an online model with only simple physics included were automatically tuned to match the output of one instrument reading (in this case the output of a transverse deflecting cavity) by using the prediction accuracy on a separate instrument as a guide (in this case a spectrometer) ~\cite{PhysRevSTAB.18.102801}. This approach can provide estimates of what the additional instrument would show non-invasively when only the more readily available diagnostics can be used, and it also only requires a simple physics model. 

New kinds of virtual instrumentation can also be created using ML. For example, artificial neural networks and support vector regression techniques have been used at LCLS to identify correlations between information obtained from slowly-updating instruments (e.g. photon energy, spectral shape of X-ray pulses) and more abundant, quickly-updating outputs (e.g. readings from beam position monitors). These correlations can be exploited to extract information that is of interest to the users about each X-ray pulse quickly and with high fidelity ~\cite{Sanchez-Gonzalez2017}.

Taking this concept a step further, the output of image-based diagnostics can also be predicted directly (thus creating a virtual version of the instrument). This has been used, for example, to predict the images produced by a destructive multi-slit transverse phase space measurement ~\cite{Edel18_1,Edel18_2,mlworkshop2018} and to predict the longitudinal phase space as displayed by a transverse deflecting cavity ~\cite{AACVD,prab_tcav}.

\subsection{Tuning, Control, and Rapid Switching Between Operating Conditions}
Particle accelerator facilities have a wide range of operational needs when it comes to tuning, optimization, and control. For example, free electron laser (FEL) user facilities have a strong incentive to reduce time spent switching between user-requested operating conditions. For other machines, users may require long-term machine stability or long-term optimization of specific beam parameters. Facilities focused on neutrino science, for example, may seek to maximize the number of protons per hour on a target.
In order to meet these demands, particle accelerators rely on databases of previous settings, on fine-tuning of machine settings by operators or online optimization routines, and dedicated feedback and control systems to maintain stable operation.
In each of these cases, ML can supplement existing techniques and procedures.

ML-based methods can reduce the need to restore old settings directly from a database. For example, ML models mapping a desired system output (such as particular beam parameters) to the required settings could cut down on the time it takes to switch between different operating conditions.
The model could also be re-trained over time, as new data accumulates and as operations change. This would shorten the time required to determine settings for some previously unseen operating state.
For an exploratory example in simulation, see ~\cite{auraleeFEL2017}, in which a neural network was trained to switch between settings needed in the injector and beamline of a compact FEL for different electron beam energies while maintaining a good match into the FEL's undulator.

ML can also be used to learn specific update rules for moving closer to an optimum. By building up representations of machine behavior, these algorithms implicitly use examples of past machine behavior to help guide the search toward optimal settings for a given operating condition.
One approach, reinforcement learning using neural networks, has been used at LCLS for tuning the undulator taper, resulting in a doubling of the power in self-seeding mode~\cite{Wu:2018tdh}.

Another approach, Bayesian optimization using Gaussian process models, has been used to tune quadrupole settings in the LCLS beamline~\cite{McIntire:2016fnl}.
Bayesian optimization can also explicitly learn to avoid disruptive combinations of settings during tuning~\cite{Krause1,Krause2}.

ML models and control policies can also be useful for systems that need to be controlled at a rate that is faster than the system settling time. 
For example, ML-based models have been trained on measured data to predict the temperature  response of a photoinjector~\cite{Edelen16,Edelen:2015lbj} and the resonant frequency response of a radio frequency quadrupole at Fermilab~\cite{Edelen:2016jgh}, both of which exhibit long-period thermal behavior but benefit from rapid control adjustments. Predictive control over the photoinjector using the model resulted in a substantially faster settling time, making it faster to switch between RF power settings while staying on-resonance.
Similarly, researchers at the Korea Atomic Energy Research Institute have created a steady-state ion source model from measured data (mapping input parameters to their final steady state output values), with the eventual aim of using it in model predictive control ~\cite{KONG201655}. Ion sources are particularly difficult to model from physics principles alone and typically exhibit highly nonlinear behavior and long response times, making them appealing targets for ML-based modeling and predictive control.

\subsection{Advanced Data Analysis}

ML methods can aid data analysis and discovery of new physics, as already seen in other fields.
Section \ref{sec:otheruses} describes computer vision algorithms that are used for high-throughput galaxy classification in astronomical surveys, or particle track identification in high energy physics detectors. At accelerator laboratories, computer vision can also be useful for processing data from image-based beam diagnostics. For example, one could automatically incorporate image data into machine models and controllers, use image data to automatically flag aberrant behavior, or automatically extract useful information for users and accelerator physicists.

Modern computational methods and ML can also be exploited to handle reconstruction tasks, such as recovering additional information about the beam from limited measurements. For example, kernel density estimation (KDE), is an unsupervised method to describe a density distribution.  KDEs were used at the Muon Ionization Cooling Experiment (MICE) to measure the phase space density of muon beams. This method enhances the precision of beam emittance measurements \cite{Mohayai:2018grg,Mohayai:2018rxn}. As another example, the field of compressed sensing incorporates prior knowledge from physical models, data, or simulations to reconstruct under-sampled targets. Researchers at UCLA and SLAC have used compressed sensing and priors trained on simulations to speed-up convergence of electron ghost imaging at the Pegasus beam line ~\cite{sli}. 

Clustering and correlation analysis can also aid exploration of complicated data sets found in accelerator simulations and measurements. Researchers at Karlsruhe Institute of Technology used K-means clustering to identify dominant modes of the microbunching instability at the Karlsruhe Research Accelerator (KARA) ring. The results led to identification of new features in the longitudinal phase space density \cite{Boltz18}.
Similarly, clustering has been used to identify different types of quenches in superconducting RF cavities at DESY ~\cite{Nawaz:IPAC2018-WEPMF058}. Such analysis can also provide guidance on promising avenues of research during early stages of larger projects.
Finally, at the Paul Scherrer Institut’s High Intensity Proton Accelerator (HIPA), researchers used correlation analysis to investigate which machine parameters contributed most to beam losses at different points along the accelerator ~\cite{mlworkshop2018}.


\clearpage
\section{Technical considerations}
Researchers should be aware of the practical requirements that must be met in order to apply ML to accelerator systems successfully. This includes, for example, the proper retention and formatting of machine data and  improving the availability and scale of computing resources (both for training and deployment). 

\subsection{Data Requirements and Availability}\label{sec:data}

The success of ML algorithms depends on the amount of available training data; specifically, the quality of a given ML solution can improve substantially with a larger numbers of training examples. Fortunately, particle accelerators are typically instrumented to accommodate this requirement. Control at many modern facilities is driven by either digital signal processors (DSPs) or field programmable gate arrays (FPGAs), and signal monitoring can occur at MHz rates or faster. Some modern facilities can produce terabytes of data per day, and this typically covers only the most relevant diagnostics information.

At most facilities, data is archived but typically only used for post-mortem analysis after failures, or for transient studies on a specific topic of interest.
In many cases there is already a sufficient amount of data in the archive to be put to use more routinely for ML applications. Much effort, however, will need to be spent in data preprocessing in order to make use of this data. Some ML techniques could also enable researchers to analyze a greater proportion of this data (for example, clustering), thus enabling it to be put to better use for improving operations and for understanding machine behavior.

\subsection{Availability and scale of computing resources}
ML-enabled data analysis is computationally intensive and often requires high-performance computing (HPC) platforms. Fortunately, these resources are becoming more accessible and can be made available for accelerator data processing. In addition, newer computing architectures, such as GPUs designed specifically for ML \footnote{\url{https://www.nvidia.com/en-us/data-center/tesla/}}, have enabled faster training of ML algorithms. GPUs can currently be purchased for just a few hundred dollars (e.g. NVIDIA GeForce GTX 1080) to a few thousand dollars (e.g. NVIDIA Tesla P100) and can easily support a small group of users. 

Furthermore, there are now many options for doing ML online through cloud-based services. For instance, Google Colaboratory \footnote{\url{https://colab.research.google.com}} allows one to use a virtual GPU through a web browser at no charge. Amazon Web Services \footnote{\url{https://aws.amazon.com/machine-learning/}} and Google Cloud \footnote{\url{https://cloud.google.com/ml-engine/}} both offer dedicated services for ML. These resources provide a convenient way of scaling up from small initial efforts without needing to purchase hardware or pay for its ongoing maintenance.

\subsection{Data formatting considerations}
Common ML scenarios in industry assume either structured data (hashed data relating, e.g., customer name and visits to a website) or homogeneous data sets such as text entered into search engines.
The relevant data for accelerator operations is mostly unstructured and heterogeneous:  floating point data and images.
Furthermore, in some cases the data may not all be logged in a uniform way. Different instruments may log data to different storage systems, or use different timestamps.

We can divide our data into two major domains: accelerator data and experiment-related data (e.g. data generated in collider detectors or light sources).
The needs of these domains, as well as the corresponding data flow and infrastructure, can greatly differ.
For many years, the HEP and light source user community has been using carefully designed data infrastructures to allow processing and analysis of data sets. 
By contrast, the accelerator domain mostly stores and handles data in a very heterogeneous and ad hoc manner.
These subsystems may store data on local disks, store data at a low rate, or not store data at all.

In order to extract unknown correlations or error signatures using ML, more comprehensive data processing, faster data transport, and increased data storage capacity will be needed.
To make accelerator data available for ML algorithms, laboratories must standardize and classify this data as early as possible in the acquisition chain.
To facilitate the use of ML frameworks, the accelerator community needs to consider classification, labeling, and storage requirements in the control system architecture, ensuring synchronicity while limiting the data rates to appropriate levels.

Much of the software requirements for making use of this data are already available and not accelerator-specific. 
Free platforms and software libraries already exist for manipulating large datasets. Furthermore, data analysis, ML, numerical, and visualization libraries in modern programming languages (e.g. Python and its associated scientific libraries, such as scikit-learn) can help to facilitate rapid development and deployment of ML algorithms.
Person-hours at accelerator laboratories can be focused on the application of these tools to specific problems.
In addition, examples of code using ML for different use cases is widely available.

\clearpage
\section{Recommendations}

\subsection{Institutional support}
The successful completion of ML projects in accelerator facilities depends on laboratory support beyond individual researchers. 
This includes support for instrumentation, control, operations, and computing staff, and their coordination.
If possible, we recommend that accelerator laboratories form a dedicated ML group.

Facilities may also need to provide additional funding and personnel for infrastructure improvements to make full use of ML.
This might include actions such as investing in or facilitating access to additional computing resources, or making improvements to data archiving systems.

Managers should not underestimate the effort needed to curate data to be used for ML. 
In addition to making use of archived data, development and testing of some algorithms (e.g. for control) will require dedicated machine time and studies. 

\subsection{Education and Intersection with Other Fields}
Mutually-beneficial collaboration between laboratories, universities, and industrial partners is essential to bring the latest developments in computer science to particle accelerators.
Laboratories should actively seek these collaborations. 

Physicists tend to already have a good grounding in probability, statistics, and data analysis.
Therefore, they already have the fundamentals in place to appreciate how ML algorithms function.
At the university level, physics students could be offered optional courses in ML, which build upon what they would have covered in probability and statistics courses.
Lectures in ML could be introduced in the various particle accelerator schools, such as the US Particle Accelerator School or the CERN Accelerator School. 
Existing education/research networks, such as the Center for Bright Beams in the US, could be leveraged to recruit graduate students to work on a common problem. 
In addition, accelerator facilities could host a local seminar series aimed at exposing accelerator physicists to current applied ML research.

\subsection{Outreach and Citizen Science}
Encouraging non-physicists to collaborate on ML problems could allow laboratories to outsource some of this work.
Examples of fora in which this could take place are online competitions such as Kaggle, in which participants are asked to provide a solution to a specific problem, with the winner being awarded a prize. As an example, a successful competition was organized in 2014 by the high energy physics community ~\cite{Adam-Bourdarios:2015pye}, and it attracted over 1700 participants.
This can also serve as a public relations and outreach exercise, as well as a way to attract students and outside researchers to work with accelerator facilities. Hackathons are an alternative way of getting accelerator physicists and ML experts together in the same room for a couple of days to work intensely on domain specific ML related problems.

\bibliography{icfa_ml_whitepaper.bib}

\begin{thebibliography}{83}%
\makeatletter
\providecommand \@ifxundefined [1]{%
 \@ifx{#1\undefined}
}%
\providecommand \@ifnum [1]{%
 \ifnum #1\expandafter \@firstoftwo
 \else \expandafter \@secondoftwo
 \fi
}%
\providecommand \@ifx [1]{%
 \ifx #1\expandafter \@firstoftwo
 \else \expandafter \@secondoftwo
 \fi
}%
\providecommand \natexlab [1]{#1}%
\providecommand \enquote  [1]{``#1''}%
\providecommand \bibnamefont  [1]{#1}%
\providecommand \bibfnamefont [1]{#1}%
\providecommand \citenamefont [1]{#1}%
\providecommand \href@noop [0]{\@secondoftwo}%
\providecommand \href [0]{\begingroup \@sanitize@url \@href}%
\providecommand \@href[1]{\@@startlink{#1}\@@href}%
\providecommand \@@href[1]{\endgroup#1\@@endlink}%
\providecommand \@sanitize@url [0]{\catcode `\\12\catcode `\$12\catcode
  `\&12\catcode `\#12\catcode `\^12\catcode `\_12\catcode `\%12\relax}%
\providecommand \@@startlink[1]{}%
\providecommand \@@endlink[0]{}%
\providecommand \url  [0]{\begingroup\@sanitize@url \@url }%
\providecommand \@url [1]{\endgroup\@href {#1}{\urlprefix }}%
\providecommand \urlprefix  [0]{URL }%
\providecommand \Eprint [0]{\href }%
\providecommand \doibase [0]{http://dx.doi.org/}%
\providecommand \selectlanguage [0]{\@gobble}%
\providecommand \bibinfo  [0]{\@secondoftwo}%
\providecommand \bibfield  [0]{\@secondoftwo}%
\providecommand \translation [1]{[#1]}%
\providecommand \BibitemOpen [0]{}%
\providecommand \bibitemStop [0]{}%
\providecommand \bibitemNoStop [0]{.\EOS\space}%
\providecommand \EOS [0]{\spacefactor3000\relax}%
\providecommand \BibitemShut  [1]{\csname bibitem#1\endcsname}%
\let\auto@bib@innerbib\@empty
\bibitem [{\citenamefont {{SLAC National Accelerator
  Laboratory}}(2018)}]{mlworkshop2018}%
  \BibitemOpen
  \bibfield  {author} {\bibinfo {author} {\bibnamefont {{SLAC National
  Accelerator Laboratory}}},\ }\\\href
  {https://conf.slac.stanford.edu/icfa-ml-2018/} {\enquote {\bibinfo {title}
  {{ICFA} beam dynamics mini-workshop: Machine learning applications for
  particle accelerators},}\ } (\bibinfo {year} {2018})\BibitemShut {NoStop}%
\bibitem [{\citenamefont {LeCun}\ \emph {et~al.}(2015)\citenamefont {LeCun},
  \citenamefont {Bengio},\ and\ \citenamefont
  {Hinton}}]{DBLP:journals/nature/LeCunBH15}%
  \BibitemOpen
  \bibfield  {author} {\bibinfo {author} {\bibfnamefont {Y.}~\bibnamefont
  {LeCun}}, \bibinfo {author} {\bibfnamefont {Y.}~\bibnamefont {Bengio}}, \
  and\ \bibinfo {author} {\bibfnamefont {G.E.}\ \bibnamefont {Hinton}},\
  }\bibfield  {title} {\enquote {\bibinfo {title} {Deep learning},}\ }\href
  {\doibase 10.1038/nature14539} {\bibfield  {journal} {\bibinfo  {journal}
  {Nature}\ }\textbf {\bibinfo {volume} {521}},\ \bibinfo {pages} {436--444}
  (\bibinfo {year} {2015})}\BibitemShut {NoStop}%
\bibitem [{\citenamefont {Awoyemi}\ \emph {et~al.}(2017)\citenamefont
  {Awoyemi}, \citenamefont {Adetunmbi},\ and\ \citenamefont
  {Oluwadare}}]{Awoyemi2017}%
  \BibitemOpen
  \bibfield  {author} {\bibinfo {author} {\bibfnamefont {J.~O.}\ \bibnamefont
  {Awoyemi}}, \bibinfo {author} {\bibfnamefont {A.~O.}\ \bibnamefont
  {Adetunmbi}}, \ and\ \bibinfo {author} {\bibfnamefont {S.~A.}\ \bibnamefont
  {Oluwadare}},\ }\bibfield  {title} {\enquote {\bibinfo {title} {Credit card
  fraud detection using machine learning techniques: A comparative analysis},}\
  }in\ \href {\doibase 10.1109/ICCNI.2017.8123782} {\emph {\bibinfo {booktitle}
  {2017 International Conference on Computing Networking and Informatics
  (ICCNI)}}}\ (\bibinfo {year} {2017})\ pp.\ \bibinfo {pages}
  {1--9}\BibitemShut {NoStop}%
\bibitem [{\citenamefont {Hirschberg}\ and\ \citenamefont
  {Manning}(2015)}]{Hirschberg:2015}%
  \BibitemOpen
  \bibfield  {author} {\bibinfo {author} {\bibfnamefont {J.}~\bibnamefont
  {Hirschberg}}\ and\ \bibinfo {author} {\bibfnamefont {C.D.}\ \bibnamefont
  {Manning}},\ }\bibfield  {title} {\enquote {\bibinfo {title} {{Advances in
  natural language processing}},}\ }\href@noop {} {\bibfield  {journal}
  {\bibinfo  {journal} {Science}\ }\textbf {\bibinfo {volume} {349}},\ \bibinfo
  {pages} {261--266} (\bibinfo {year} {2015})}\BibitemShut {NoStop}%
\bibitem [{\citenamefont {{Daresbury
  Laboratory}}(2018)}]{daresburyworkshop2018}%
  \BibitemOpen
  \bibfield  {author} {\bibinfo {author} {\bibnamefont {{Daresbury
  Laboratory}}},\ }\href {https://www.cockcroft.ac.uk/events/ICPA/} {\enquote
  {\bibinfo {title} {Intelligent controls for particle accelerators
  workshop},}\ } (\bibinfo {year} {2018})\BibitemShut {NoStop}%
\bibitem [{\citenamefont {Edelen}\ \emph
  {et~al.}(2016{\natexlab{a}})\citenamefont {Edelen}, \citenamefont {Biedron},
  \citenamefont {Chase}, \citenamefont {Edstrom}, \citenamefont {Milton},\ and\
  \citenamefont {Stabile}}]{Edelen16}%
  \BibitemOpen
  \bibfield  {author} {\bibinfo {author} {\bibfnamefont {A.~L.}\ \bibnamefont
  {Edelen}}, \bibinfo {author} {\bibfnamefont {S.~G.}\ \bibnamefont {Biedron}},
  \bibinfo {author} {\bibfnamefont {B.~E.}\ \bibnamefont {Chase}}, \bibinfo
  {author} {\bibfnamefont {D.}~\bibnamefont {Edstrom}}, \bibinfo {author}
  {\bibfnamefont {S.~V.}\ \bibnamefont {Milton}}, \ and\ \bibinfo {author}
  {\bibfnamefont {P.}~\bibnamefont {Stabile}},\ }\bibfield  {title} {\enquote
  {\bibinfo {title} {{Neural Networks for Modeling and Control of Particle
  Accelerators}},}\ }\href {\doibase 10.1109/TNS.2016.2543203} {\bibfield
  {journal} {\bibinfo  {journal} {IEEE Trans. Nucl. Sci.}\ }\textbf {\bibinfo
  {volume} {63}},\ \bibinfo {pages} {878--897} (\bibinfo {year}
  {2016}{\natexlab{a}})},\ \Eprint {http://arxiv.org/abs/1610.06151}
  {arXiv:1610.06151 [physics.acc-ph]} \BibitemShut {NoStop}%
\bibitem [{\citenamefont {Schmidhuber}(2015)}]{SCHMIDHUBER201585}%
  \BibitemOpen
  \bibfield  {author} {\bibinfo {author} {\bibfnamefont {J.}~\bibnamefont
  {Schmidhuber}},\ }\bibfield  {title} {\enquote {\bibinfo {title} {Deep
  learning in neural networks: An overview},}\ }\href {\doibase
  https://doi.org/10.1016/j.neunet.2014.09.003} {\bibfield  {journal} {\bibinfo
   {journal} {Neural Networks}\ }\textbf {\bibinfo {volume} {61}},\ \bibinfo
  {pages} {85 -- 117} (\bibinfo {year} {2015})}\BibitemShut {NoStop}%
\bibitem [{\citenamefont {Ball}\ and\ \citenamefont
  {Brunner}(2010)}]{doi:10.1142/S0218271810017160}%
  \BibitemOpen
  \bibfield  {author} {\bibinfo {author} {\bibfnamefont {N.M.}\ \bibnamefont
  {Ball}}\ and\ \bibinfo {author} {\bibfnamefont {R.J.}\ \bibnamefont
  {Brunner}},\ }\bibfield  {title} {\enquote {\bibinfo {title} {Data mining and
  machine learning in astronomy},}\ }\href {\doibase 10.1142/S0218271810017160}
  {\bibfield  {journal} {\bibinfo  {journal} {International Journal of Modern
  Physics D}\ }\textbf {\bibinfo {volume} {19}},\ \bibinfo {pages} {1049--1106}
  (\bibinfo {year} {2010})},\ \Eprint
  {http://arxiv.org/abs/https://www.worldscientific.com/doi/pdf/10.1142/S0218271810017160}
  {https://www.worldscientific.com/doi/pdf/10.1142/S0218271810017160}
  \BibitemShut {NoStop}%
\bibitem [{\citenamefont {Way}\ \emph {et~al.}(2012)\citenamefont {Way},
  \citenamefont {Scargle}, \citenamefont {Ali},\ and\ \citenamefont
  {Srivastava}}]{way2017}%
  \BibitemOpen
  \bibinfo {editor} {\bibfnamefont {M.}~\bibnamefont {Way}}, \bibinfo {editor}
  {\bibfnamefont {J.}~\bibnamefont {Scargle}}, \bibinfo {editor} {\bibfnamefont
  {K.}~\bibnamefont {Ali}}, \ and\ \bibinfo {editor} {\bibfnamefont
  {A.}~\bibnamefont {Srivastava}},\ eds.,\ \href@noop {} {\emph {\bibinfo
  {title} {Advances in Machine Learning and Data Mining for Astronomy}}}\
  (\bibinfo  {publisher} {New York: Chapman and Hall/CRC},\ \bibinfo {year}
  {2012})\BibitemShut {NoStop}%
\bibitem [{\citenamefont {Graff}\ \emph {et~al.}(2013)\citenamefont {Graff},
  \citenamefont {Feroz}, \citenamefont {Hobson},\ and\ \citenamefont
  {Lasenby}}]{Graff2013}%
  \BibitemOpen
  \bibfield  {author} {\bibinfo {author} {\bibfnamefont {P.}~\bibnamefont
  {Graff}}, \bibinfo {author} {\bibfnamefont {F.}~\bibnamefont {Feroz}},
  \bibinfo {author} {\bibfnamefont {M.~P.}\ \bibnamefont {Hobson}}, \ and\
  \bibinfo {author} {\bibfnamefont {A.}~\bibnamefont {Lasenby}},\ }\bibfield
  {title} {\enquote {\bibinfo {title} {Neural networks for astronomical data
  analysis and bayesian inference},}\ }in\ \href {\doibase
  10.1109/ICDMW.2013.82} {\emph {\bibinfo {booktitle} {2013 IEEE 13th
  International Conference on Data Mining Workshops}}}\ (\bibinfo {year}
  {2013})\ pp.\ \bibinfo {pages} {16--23}\BibitemShut {NoStop}%
\bibitem [{\citenamefont {Graff}\ \emph {et~al.}(2014)\citenamefont {Graff},
  \citenamefont {Feroz}, \citenamefont {Hobson},\ and\ \citenamefont
  {Lasenby}}]{doi:10.1093/mnras/stu642}%
  \BibitemOpen
  \bibfield  {author} {\bibinfo {author} {\bibfnamefont {P.}~\bibnamefont
  {Graff}}, \bibinfo {author} {\bibfnamefont {F.}~\bibnamefont {Feroz}},
  \bibinfo {author} {\bibfnamefont {M.P.}\ \bibnamefont {Hobson}}, \ and\
  \bibinfo {author} {\bibfnamefont {A.}~\bibnamefont {Lasenby}},\ }\bibfield
  {title} {\enquote {\bibinfo {title} {Skynet: an efficient and robust neural
  network training tool for machine learning in astronomy},}\ }\href {\doibase
  10.1093/mnras/stu642} {\bibfield  {journal} {\bibinfo  {journal} {Monthly
  Notices of the Royal Astronomical Society}\ }\textbf {\bibinfo {volume}
  {441}},\ \bibinfo {pages} {1741--1759} (\bibinfo {year} {2014})}\BibitemShut
  {NoStop}%
\bibitem [{\citenamefont {Biswas}\ \emph {et~al.}(2013)\citenamefont {Biswas},
  \citenamefont {Blackburn}, \citenamefont {Cao}, \citenamefont {Essick},
  \citenamefont {Hodge}, \citenamefont {Katsavounidis}, \citenamefont {Kim},
  \citenamefont {Kim}, \citenamefont {Le~Bigot}, \citenamefont {Lee},
  \citenamefont {Oh} \emph {et~al.}}]{PhysRevD.88.062003}%
  \BibitemOpen
  \bibfield  {author} {\bibinfo {author} {\bibfnamefont {R.}~\bibnamefont
  {Biswas}}, \bibinfo {author} {\bibfnamefont {L.}~\bibnamefont {Blackburn}},
  \bibinfo {author} {\bibfnamefont {J.}~\bibnamefont {Cao}}, \bibinfo {author}
  {\bibfnamefont {R.}~\bibnamefont {Essick}}, \bibinfo {author} {\bibfnamefont
  {K.A.}\ \bibnamefont {Hodge}}, \bibinfo {author} {\bibfnamefont
  {E.}~\bibnamefont {Katsavounidis}}, \bibinfo {author} {\bibfnamefont
  {K.}~\bibnamefont {Kim}}, \bibinfo {author} {\bibfnamefont {Y.-M.}\
  \bibnamefont {Kim}}, \bibinfo {author} {\bibfnamefont {E.-O.}\ \bibnamefont
  {Le~Bigot}}, \bibinfo {author} {\bibfnamefont {C.-H.}\ \bibnamefont {Lee}},
  \bibinfo {author} {\bibfnamefont {J.J.}\ \bibnamefont {Oh}},  \emph
  {et~al.},\ }\bibfield  {title} {\enquote {\bibinfo {title} {Application of
  machine learning algorithms to the study of noise artifacts in
  gravitational-wave data},}\ }\href {\doibase 10.1103/PhysRevD.88.062003}
  {\bibfield  {journal} {\bibinfo  {journal} {Phys. Rev. D}\ }\textbf {\bibinfo
  {volume} {88}},\ \bibinfo {pages} {062003} (\bibinfo {year}
  {2013})}\BibitemShut {NoStop}%
\bibitem [{\citenamefont {Bhat}(2011{\natexlab{a}})}]{pbhat}%
  \BibitemOpen
  \bibfield  {author} {\bibinfo {author} {\bibfnamefont {P.}~\bibnamefont
  {Bhat}},\ }\bibfield  {title} {\enquote {\bibinfo {title} {Multivariate
  analysis methods in particle physics},}\ }\href@noop {} {\bibfield  {journal}
  {\bibinfo  {journal} {Annual Review of Nuclear and Particle Science}\
  }\textbf {\bibinfo {volume} {121}},\ \bibinfo {pages} {281--309} (\bibinfo
  {year} {2011}{\natexlab{a}})}\BibitemShut {NoStop}%
\bibitem [{\citenamefont {Aurisanoa}\ \emph {et~al.}(2016)\citenamefont
  {Aurisanoa}, \citenamefont {Radovic}, \citenamefont {Rocco}, \citenamefont
  {Himmel}, \citenamefont {Messier}, \citenamefont {Niner}, \citenamefont
  {Pawloski}, \citenamefont {Psihas}, \citenamefont {Sousa},\ and\
  \citenamefont {Vahle}}]{AAurisano:2016}%
  \BibitemOpen
  \bibfield  {author} {\bibinfo {author} {\bibfnamefont {A.}~\bibnamefont
  {Aurisanoa}}, \bibinfo {author} {\bibfnamefont {A.}~\bibnamefont {Radovic}},
  \bibinfo {author} {\bibfnamefont {D.}~\bibnamefont {Rocco}}, \bibinfo
  {author} {\bibfnamefont {A.}~\bibnamefont {Himmel}}, \bibinfo {author}
  {\bibfnamefont {M.D.}\ \bibnamefont {Messier}}, \bibinfo {author}
  {\bibfnamefont {E.}~\bibnamefont {Niner}}, \bibinfo {author} {\bibfnamefont
  {G.}~\bibnamefont {Pawloski}}, \bibinfo {author} {\bibfnamefont
  {F.}~\bibnamefont {Psihas}}, \bibinfo {author} {\bibfnamefont
  {A.}~\bibnamefont {Sousa}}, \ and\ \bibinfo {author} {\bibfnamefont
  {P.}~\bibnamefont {Vahle}},\ }\bibfield  {title} {\enquote {\bibinfo {title}
  {{A convolutional neural network neutrino event classifier}},}\ }\href
  {\doibase https://doi.org/10.1088/1748-0221/11/09/P09001} {\bibfield
  {journal} {\bibinfo  {journal} {Journal of Instrumentation}\ }\textbf
  {\bibinfo {volume} {11}} (\bibinfo {year} {2016}),\
  https://doi.org/10.1088/1748-0221/11/09/P09001}\BibitemShut {NoStop}%
\bibitem [{\citenamefont {Baldi}\ \emph {et~al.}(2014)\citenamefont {Baldi},
  \citenamefont {Sadowski},\ and\ \citenamefont
  {Whiteson}}]{Baldi2014SearchingFE}%
  \BibitemOpen
  \bibfield  {author} {\bibinfo {author} {\bibfnamefont {P.}~\bibnamefont
  {Baldi}}, \bibinfo {author} {\bibfnamefont {Paul~D.}\ \bibnamefont
  {Sadowski}}, \ and\ \bibinfo {author} {\bibfnamefont {Daniel}\ \bibnamefont
  {Whiteson}},\ }\bibfield  {title} {\enquote {\bibinfo {title} {Searching for
  exotic particles in high-energy physics with deep learning.}}\ }\href@noop {}
  {\bibfield  {journal} {\bibinfo  {journal} {Nature communications}\ }\textbf
  {\bibinfo {volume} {5}},\ \bibinfo {pages} {4308} (\bibinfo {year}
  {2014})}\BibitemShut {NoStop}%
\bibitem [{\citenamefont
  {Bhat}(2011{\natexlab{b}})}]{doi:10.1146/annurev.nucl.012809.104427}%
  \BibitemOpen
  \bibfield  {author} {\bibinfo {author} {\bibfnamefont {P.C.}\ \bibnamefont
  {Bhat}},\ }\bibfield  {title} {\enquote {\bibinfo {title} {Multivariate
  analysis methods in particle physics},}\ }\href {\doibase
  10.1146/annurev.nucl.012809.104427} {\bibfield  {journal} {\bibinfo
  {journal} {Annual Review of Nuclear and Particle Science}\ }\textbf {\bibinfo
  {volume} {61}},\ \bibinfo {pages} {281--309} (\bibinfo {year}
  {2011}{\natexlab{b}})},\ \Eprint
  {http://arxiv.org/abs/https://doi.org/10.1146/annurev.nucl.012809.104427}
  {https://doi.org/10.1146/annurev.nucl.012809.104427} \BibitemShut {NoStop}%
\bibitem [{\citenamefont {Whiteson}\ and\ \citenamefont
  {Whiteson}(2009)}]{WHITESON20091203}%
  \BibitemOpen
  \bibfield  {author} {\bibinfo {author} {\bibfnamefont {S.}~\bibnamefont
  {Whiteson}}\ and\ \bibinfo {author} {\bibfnamefont {D.}~\bibnamefont
  {Whiteson}},\ }\bibfield  {title} {\enquote {\bibinfo {title} {Machine
  learning for event selection in high energy physics},}\ }\href {\doibase
  https://doi.org/10.1016/j.engappai.2009.05.004} {\bibfield  {journal}
  {\bibinfo  {journal} {Engineering Applications of Artificial Intelligence}\
  }\textbf {\bibinfo {volume} {22}},\ \bibinfo {pages} {1203 -- 1217} (\bibinfo
  {year} {2009})}\BibitemShut {NoStop}%
\bibitem [{\citenamefont {Radovic}\ \emph {et~al.}(2018)\citenamefont
  {Radovic}, \citenamefont {Williams}, \citenamefont {Rousseau}, \citenamefont
  {Kagan}, \citenamefont {Bonacorsi}, \citenamefont {Himmel}, \citenamefont
  {Aurisano}, \citenamefont {Terao},\ and\ \citenamefont
  {Wongjirad}}]{Radovic}%
  \BibitemOpen
  \bibfield  {author} {\bibinfo {author} {\bibfnamefont {A.}~\bibnamefont
  {Radovic}}, \bibinfo {author} {\bibfnamefont {M.}~\bibnamefont {Williams}},
  \bibinfo {author} {\bibfnamefont {D.}~\bibnamefont {Rousseau}}, \bibinfo
  {author} {\bibfnamefont {M.}~\bibnamefont {Kagan}}, \bibinfo {author}
  {\bibfnamefont {D.}~\bibnamefont {Bonacorsi}}, \bibinfo {author}
  {\bibfnamefont {A.}~\bibnamefont {Himmel}}, \bibinfo {author} {\bibfnamefont
  {A.}~\bibnamefont {Aurisano}}, \bibinfo {author} {\bibfnamefont
  {K.}~\bibnamefont {Terao}}, \ and\ \bibinfo {author} {\bibfnamefont
  {T.}~\bibnamefont {Wongjirad}},\ }\bibfield  {title} {\enquote {\bibinfo
  {title} {Machine learning at the energy and intensity frontiers of particle
  physics},}\ }\href {\doibase 10.1038/s41586-018-0361-2} {\bibfield  {journal}
  {\bibinfo  {journal} {Nature}\ }\textbf {\bibinfo {volume} {560}},\ \bibinfo
  {pages} {41--48} (\bibinfo {year} {2018})}\BibitemShut {NoStop}%
\bibitem [{\citenamefont {Adam-Bourdarios}\ \emph {et~al.}(2015)\citenamefont
  {Adam-Bourdarios}, \citenamefont {Cowan}, \citenamefont {Germain-Renaud},
  \citenamefont {Guyon}, \citenamefont {Kégl},\ and\ \citenamefont
  {Rousseau}}]{Adam-Bourdarios:2015pye}%
  \BibitemOpen
  \bibfield  {author} {\bibinfo {author} {\bibfnamefont {C.}~\bibnamefont
  {Adam-Bourdarios}}, \bibinfo {author} {\bibfnamefont {G.}~\bibnamefont
  {Cowan}}, \bibinfo {author} {\bibfnamefont {C.}~\bibnamefont
  {Germain-Renaud}}, \bibinfo {author} {\bibfnamefont {I.}~\bibnamefont
  {Guyon}}, \bibinfo {author} {\bibfnamefont {B.}~\bibnamefont {Kégl}}, \ and\
  \bibinfo {author} {\bibfnamefont {D.}~\bibnamefont {Rousseau}},\ }\bibfield
  {title} {\enquote {\bibinfo {title} {{The Higgs Machine Learning
  Challenge}},}\ }\bibfield  {booktitle} {\emph {\bibinfo {booktitle}
  {{Proceedings, 21st International Conference on Computing in High Energy and
  Nuclear Physics (CHEP 2015): Okinawa, Japan, April 13-17, 2015}}},\ }\href
  {\doibase 10.1088/1742-6596/664/7/072015} {\bibfield  {journal} {\bibinfo
  {journal} {J. Phys. Conf. Ser.}\ }\textbf {\bibinfo {volume} {664}},\
  \bibinfo {pages} {072015} (\bibinfo {year} {2015})}\BibitemShut {NoStop}%
\bibitem [{\citenamefont {Esteva}\ \emph {et~al.}(2017)\citenamefont {Esteva},
  \citenamefont {Kuprel}, \citenamefont {Novoa}, \citenamefont {Ko},
  \citenamefont {Swetter}, \citenamefont {Blau},\ and\ \citenamefont
  {Thrun}}]{DBLP:journals/nature/EstevaKNKSBT17}%
  \BibitemOpen
  \bibfield  {author} {\bibinfo {author} {\bibfnamefont {A.}~\bibnamefont
  {Esteva}}, \bibinfo {author} {\bibfnamefont {B.}~\bibnamefont {Kuprel}},
  \bibinfo {author} {\bibfnamefont {R.A.}\ \bibnamefont {Novoa}}, \bibinfo
  {author} {\bibfnamefont {J.}~\bibnamefont {Ko}}, \bibinfo {author}
  {\bibfnamefont {S.M.}\ \bibnamefont {Swetter}}, \bibinfo {author}
  {\bibfnamefont {H.M.}\ \bibnamefont {Blau}}, \ and\ \bibinfo {author}
  {\bibfnamefont {S.}~\bibnamefont {Thrun}},\ }\bibfield  {title} {\enquote
  {\bibinfo {title} {Dermatologist-level classification of skin cancer with
  deep neural networks},}\ }\href {\doibase 10.1038/nature21056} {\bibfield
  {journal} {\bibinfo  {journal} {Nature}\ }\textbf {\bibinfo {volume} {542}},\
  \bibinfo {pages} {115--118} (\bibinfo {year} {2017})}\BibitemShut {NoStop}%
\bibitem [{\citenamefont {Zhang}\ and\ \citenamefont {Liu}(2004)}]{Zhang04}%
  \BibitemOpen
  \bibfield  {author} {\bibinfo {author} {\bibfnamefont {J.}~\bibnamefont
  {Zhang}}\ and\ \bibinfo {author} {\bibfnamefont {Y.}~\bibnamefont {Liu}},\
  }\bibfield  {title} {\enquote {\bibinfo {title} {Cervical cancer detection
  using svm based feature screening},}\ }in\ \href@noop {} {\emph {\bibinfo
  {booktitle} {International Conference on Medical Image Computing and
  Computer-Assisted Intervention}}}\ (\bibinfo {organization} {Springer},\
  \bibinfo {year} {2004})\ pp.\ \bibinfo {pages} {873--880}\BibitemShut
  {NoStop}%
\bibitem [{\citenamefont {{Rajpurkar}}\ \emph {et~al.}(2017)\citenamefont
  {{Rajpurkar}}, \citenamefont {{Hannun}}, \citenamefont {{Haghpanahi}},
  \citenamefont {{Bourn}},\ and\ \citenamefont {{Ng}}}]{2017arXiv170701836R}%
  \BibitemOpen
  \bibfield  {author} {\bibinfo {author} {\bibfnamefont {P.}~\bibnamefont
  {{Rajpurkar}}}, \bibinfo {author} {\bibfnamefont {A.~Y.}\ \bibnamefont
  {{Hannun}}}, \bibinfo {author} {\bibfnamefont {M.}~\bibnamefont
  {{Haghpanahi}}}, \bibinfo {author} {\bibfnamefont {C.}~\bibnamefont
  {{Bourn}}}, \ and\ \bibinfo {author} {\bibfnamefont {A.~Y.}\ \bibnamefont
  {{Ng}}},\ }\bibfield  {title} {\enquote {\bibinfo {title}
  {{Cardiologist-Level Arrhythmia Detection with Convolutional Neural
  Networks}},}\ }\href@noop {} {\bibfield  {journal} {\bibinfo  {journal}
  {ArXiv e-prints}\ } (\bibinfo {year} {2017})},\ \Eprint
  {http://arxiv.org/abs/1707.01836} {arXiv:1707.01836 [cs.CV]} \BibitemShut
  {NoStop}%
\bibitem [{\citenamefont {Deisenroth}\ \emph {et~al.}(2013)\citenamefont
  {Deisenroth}, \citenamefont {Neumann},\ and\ \citenamefont
  {Peters}}]{ROB-021}%
  \BibitemOpen
  \bibfield  {author} {\bibinfo {author} {\bibfnamefont {M.P.}\ \bibnamefont
  {Deisenroth}}, \bibinfo {author} {\bibfnamefont {G.}~\bibnamefont {Neumann}},
  \ and\ \bibinfo {author} {\bibfnamefont {J.}~\bibnamefont {Peters}},\
  }\bibfield  {title} {\enquote {\bibinfo {title} {A survey on policy search
  for robotics},}\ }\href@noop {} {\bibfield  {journal} {\bibinfo  {journal}
  {Foundations and Trends in Robotics}\ }\textbf {\bibinfo {volume} {2}},\
  \bibinfo {pages} {1--142} (\bibinfo {year} {2013})}\BibitemShut {NoStop}%
\bibitem [{\citenamefont {Srihari}\ and\ \citenamefont
  {Kuebert}(1997)}]{Srihari1997}%
  \BibitemOpen
  \bibfield  {author} {\bibinfo {author} {\bibfnamefont {S.~N.}\ \bibnamefont
  {Srihari}}\ and\ \bibinfo {author} {\bibfnamefont {E.~J.}\ \bibnamefont
  {Kuebert}},\ }\bibfield  {title} {\enquote {\bibinfo {title} {Integration of
  hand-written address interpretation technology into the united states postal
  service remote computer reader system},}\ }in\ \href {\doibase
  10.1109/ICDAR.1997.620640} {\emph {\bibinfo {booktitle} {Proceedings of the
  Fourth International Conference on Document Analysis and Recognition}}},\
  Vol.~\bibinfo {volume} {2}\ (\bibinfo {year} {1997})\ pp.\ \bibinfo {pages}
  {892--896 vol.2}\BibitemShut {NoStop}%
\bibitem [{\citenamefont {Murari}\ \emph
  {et~al.}(2012{\natexlab{a}})\citenamefont {Murari}, \citenamefont {Mazon},
  \citenamefont {Martin}, \citenamefont {Vagliasindi},\ and\ \citenamefont
  {Gelfusa}}]{Murari2012}%
  \BibitemOpen
  \bibfield  {author} {\bibinfo {author} {\bibfnamefont {A.}~\bibnamefont
  {Murari}}, \bibinfo {author} {\bibfnamefont {D.}~\bibnamefont {Mazon}},
  \bibinfo {author} {\bibfnamefont {N.}~\bibnamefont {Martin}}, \bibinfo
  {author} {\bibfnamefont {G.}~\bibnamefont {Vagliasindi}}, \ and\ \bibinfo
  {author} {\bibfnamefont {M.}~\bibnamefont {Gelfusa}},\ }\bibfield  {title}
  {\enquote {\bibinfo {title} {Exploratory data analysis techniques to
  determine the dimensionality of complex nonlinear phenomena: The l-to-h
  transition at jet as a case study},}\ }\href {\doibase
  10.1109/TPS.2012.2187682} {\bibfield  {journal} {\bibinfo  {journal} {IEEE
  Transactions on Plasma Science}\ }\textbf {\bibinfo {volume} {40}},\ \bibinfo
  {pages} {1386--1394} (\bibinfo {year} {2012}{\natexlab{a}})}\BibitemShut
  {NoStop}%
\bibitem [{\citenamefont {Murari}\ \emph
  {et~al.}(2012{\natexlab{b}})\citenamefont {Murari}, \citenamefont
  {Buscarino}, \citenamefont {Fortuna}, \citenamefont {Frasca}, \citenamefont
  {Iachello},\ and\ \citenamefont {Mazzitelli}}]{Murari2012IdentifyingJI}%
  \BibitemOpen
  \bibfield  {author} {\bibinfo {author} {\bibfnamefont {A.}~\bibnamefont
  {Murari}}, \bibinfo {author} {\bibfnamefont {A.}~\bibnamefont {Buscarino}},
  \bibinfo {author} {\bibfnamefont {L.}~\bibnamefont {Fortuna}}, \bibinfo
  {author} {\bibfnamefont {M.}~\bibnamefont {Frasca}}, \bibinfo {author}
  {\bibfnamefont {M.}~\bibnamefont {Iachello}}, \ and\ \bibinfo {author}
  {\bibfnamefont {G.}~\bibnamefont {Mazzitelli}},\ }\bibfield  {title}
  {\enquote {\bibinfo {title} {Identifying jet instabilities with neural
  networks},}\ }\href@noop {} {\bibfield  {journal} {\bibinfo  {journal} {2012
  16th IEEE Mediterranean Electrotechnical Conference}\ ,\ \bibinfo {pages}
  {932--935}} (\bibinfo {year} {2012}{\natexlab{b}})}\BibitemShut {NoStop}%
\bibitem [{\citenamefont {Murari}\ \emph {et~al.}(2013)\citenamefont {Murari},
  \citenamefont {Arena}, \citenamefont {Buscarino}, \citenamefont {Fortuna},\
  and\ \citenamefont {Iachello}}]{MURARI20132}%
  \BibitemOpen
  \bibfield  {author} {\bibinfo {author} {\bibfnamefont {A.}~\bibnamefont
  {Murari}}, \bibinfo {author} {\bibfnamefont {P.}~\bibnamefont {Arena}},
  \bibinfo {author} {\bibfnamefont {A.}~\bibnamefont {Buscarino}}, \bibinfo
  {author} {\bibfnamefont {L.}~\bibnamefont {Fortuna}}, \ and\ \bibinfo
  {author} {\bibfnamefont {M.}~\bibnamefont {Iachello}},\ }\bibfield  {title}
  {\enquote {\bibinfo {title} {On the identification of instabilities with
  neural networks on jet},}\ }\href {\doibase
  https://doi.org/10.1016/j.nima.2013.03.039} {\bibfield  {journal} {\bibinfo
  {journal} {Nucl. Instrum. Methods Phys. Res. A}\ }\textbf
  {\bibinfo {volume} {720}},\ \bibinfo {pages} {2 -- 6} (\bibinfo {year}
  {2013})},\ \bibinfo {note} {selected papers from the 2nd International
  Conference Frontiers in Diagnostic Technologies (ICFDT2)}\BibitemShut
  {NoStop}%
\bibitem [{\citenamefont {Baltz}\ \emph {et~al.}(2017)\citenamefont {Baltz},
  \citenamefont {Trask}, \citenamefont {Binderbauer}, \citenamefont {Dikovsky},
  \citenamefont {Gota}, \citenamefont {Mendoza}, \citenamefont {Platt},\ and\
  \citenamefont {Riley}}]{optometrist-algorithm}%
  \BibitemOpen
  \bibfield  {author} {\bibinfo {author} {\bibfnamefont {E.~A.}\ \bibnamefont
  {Baltz}}, \bibinfo {author} {\bibfnamefont {E.}~\bibnamefont {Trask}},
  \bibinfo {author} {\bibfnamefont {M.}~\bibnamefont {Binderbauer}}, \bibinfo
  {author} {\bibfnamefont {M.}~\bibnamefont {Dikovsky}}, \bibinfo {author}
  {\bibfnamefont {H.}~\bibnamefont {Gota}}, \bibinfo {author} {\bibfnamefont
  {R.}~\bibnamefont {Mendoza}}, \bibinfo {author} {\bibfnamefont {J.~C.}\
  \bibnamefont {Platt}}, \ and\ \bibinfo {author} {\bibfnamefont {P.~F.}\
  \bibnamefont {Riley}},\ }\bibfield  {title} {\enquote {\bibinfo {title}
  {Achievement of sustained net plasma heating in a fusion experiment with the
  optometrist algorithm},}\ }\href {\doibase 10.1038/s41598-017-06645-7}
  {\bibfield  {journal} {\bibinfo  {journal} {Scientific Reports}\ }\textbf
  {\bibinfo {volume} {7}},\ \bibinfo {pages} {6425} (\bibinfo {year}
  {2017})}\BibitemShut {NoStop}%
\bibitem [{\citenamefont {Higo}\ \emph {et~al.}(1986)\citenamefont {Higo},
  \citenamefont {Shoaee},\ and\ \citenamefont {Spencer}}]{Higo86}%
  \BibitemOpen
  \bibfield  {author} {\bibinfo {author} {\bibfnamefont {T.}~\bibnamefont
  {Higo}}, \bibinfo {author} {\bibfnamefont {H.}~\bibnamefont {Shoaee}}, \ and\
  \bibinfo {author} {\bibfnamefont {J}~\bibnamefont {Spencer}},\ }\bibfield
  {title} {\enquote {\bibinfo {title} {Some applications of ai to the problems
  of accelerator physics},}\ }in\ \href@noop {} {\emph {\bibinfo {booktitle}
  {Conf. Proc.}}},\ Vol.\ \bibinfo {volume} {870316}\ (\bibinfo {year} {1986})\
  p.\ \bibinfo {pages} {701}\BibitemShut {NoStop}%
\bibitem [{\citenamefont {Weygand}(1987)}]{Weygand87}%
  \BibitemOpen
  \bibfield  {author} {\bibinfo {author} {\bibfnamefont {D.P.}\ \bibnamefont
  {Weygand}},\ }\href@noop {} {\emph {\bibinfo {title} {Artificial intelligence
  and accelerator control}}},\ \bibinfo {type} {Tech. Rep.}\ (\bibinfo
  {institution} {Brookhaven National Lab., Upton, NY (USA)},\ \bibinfo {year}
  {1987})\BibitemShut {NoStop}%
\bibitem [{\citenamefont {Skarek}\ and\ \citenamefont
  {Varga}(1996)}]{Skarek96}%
  \BibitemOpen
  \bibfield  {author} {\bibinfo {author} {\bibfnamefont {P.}~\bibnamefont
  {Skarek}}\ and\ \bibinfo {author} {\bibfnamefont {L.}~\bibnamefont {Varga}},\
  }\bibfield  {title} {\enquote {\bibinfo {title} {Multi-agent cooperation for
  particle accelerator control},}\ }\href@noop {} {\bibfield  {journal}
  {\bibinfo  {journal} {Expert Systems with applications}\ }\textbf {\bibinfo
  {volume} {11}},\ \bibinfo {pages} {481--487} (\bibinfo {year}
  {1996})}\BibitemShut {NoStop}%
\bibitem [{\citenamefont {Schultz}\ and\ \citenamefont
  {Brown}(1990)}]{Schultz90}%
  \BibitemOpen
  \bibfield  {author} {\bibinfo {author} {\bibfnamefont {D.E.}\ \bibnamefont
  {Schultz}}\ and\ \bibinfo {author} {\bibfnamefont {P.A.}\ \bibnamefont
  {Brown}},\ }\bibfield  {title} {\enquote {\bibinfo {title} {The development
  of an expert system to tune a beam line},}\ }\href@noop {} {\bibfield
  {journal} {\bibinfo  {journal} {Nuclear Instruments and Methods in Physics
  Research Section A: Accelerators, Spectrometers, Detectors and Associated
  Equipment}\ }\textbf {\bibinfo {volume} {293}},\ \bibinfo {pages} {486--490}
  (\bibinfo {year} {1990})}\BibitemShut {NoStop}%
\bibitem [{\citenamefont {Fiesler}\ and\ \citenamefont
  {Campbell}(1999)}]{Fiesler99}%
  \BibitemOpen
  \bibfield  {author} {\bibinfo {author} {\bibfnamefont {E.}~\bibnamefont
  {Fiesler}}\ and\ \bibinfo {author} {\bibfnamefont {S.R.}\ \bibnamefont
  {Campbell}},\ }\bibfield  {title} {\enquote {\bibinfo {title} {Hybrid neural
  networks and their application to particle accelerator control},}\ }in\
  \href@noop {} {\emph {\bibinfo {booktitle} {Applications and Science of
  Neural Networks, Fuzzy Systems, and Evolutionary Computation II}}},\ Vol.\
  \bibinfo {volume} {3812}\ (\bibinfo {organization} {International Society for
  Optics and Photonics},\ \bibinfo {year} {1999})\ pp.\ \bibinfo {pages}
  {132--143}\BibitemShut {NoStop}%
\bibitem [{\citenamefont {Howell}\ \emph {et~al.}(1990)\citenamefont {Howell},
  \citenamefont {Barnes}, \citenamefont {Brown}, \citenamefont {Flake},
  \citenamefont {Jones}, \citenamefont {Lee}, \citenamefont {Qian},\ and\
  \citenamefont {Wright}}]{Howell90}%
  \BibitemOpen
  \bibfield  {author} {\bibinfo {author} {\bibfnamefont {J.A.}\ \bibnamefont
  {Howell}}, \bibinfo {author} {\bibfnamefont {C.W.}\ \bibnamefont {Barnes}},
  \bibinfo {author} {\bibfnamefont {S.K.}\ \bibnamefont {Brown}}, \bibinfo
  {author} {\bibfnamefont {G.W.}\ \bibnamefont {Flake}}, \bibinfo {author}
  {\bibfnamefont {R.D.}\ \bibnamefont {Jones}}, \bibinfo {author}
  {\bibfnamefont {Y.C.}\ \bibnamefont {Lee}}, \bibinfo {author} {\bibfnamefont
  {S.}~\bibnamefont {Qian}}, \ and\ \bibinfo {author} {\bibfnamefont {R.M.}\
  \bibnamefont {Wright}},\ }\bibfield  {title} {\enquote {\bibinfo {title}
  {Control of a negative-ion accelerator source using neural networks},}\
  }\href@noop {} {\bibfield  {journal} {\bibinfo  {journal} {Nuclear
  Instruments and Methods in Physics Research Section A: Accelerators,
  Spectrometers, Detectors and Associated Equipment}\ }\textbf {\bibinfo
  {volume} {293}},\ \bibinfo {pages} {517--522} (\bibinfo {year}
  {1990})}\BibitemShut {NoStop}%
\bibitem [{\citenamefont {Mead}\ \emph {et~al.}(1992)\citenamefont {Mead},
  \citenamefont {Bowling}, \citenamefont {Brown}, \citenamefont {Jones},
  \citenamefont {Barnes}, \citenamefont {Gibson}, \citenamefont {Goulding},\
  and\ \citenamefont {Lee}}]{Mead92}%
  \BibitemOpen
  \bibfield  {author} {\bibinfo {author} {\bibfnamefont {W.C.}\ \bibnamefont
  {Mead}}, \bibinfo {author} {\bibfnamefont {P.S.}\ \bibnamefont {Bowling}},
  \bibinfo {author} {\bibfnamefont {S.K.}\ \bibnamefont {Brown}}, \bibinfo
  {author} {\bibfnamefont {R.D.}\ \bibnamefont {Jones}}, \bibinfo {author}
  {\bibfnamefont {C.W.}\ \bibnamefont {Barnes}}, \bibinfo {author}
  {\bibfnamefont {H.E.}\ \bibnamefont {Gibson}}, \bibinfo {author}
  {\bibfnamefont {J.R.}\ \bibnamefont {Goulding}}, \ and\ \bibinfo {author}
  {\bibfnamefont {Y.C.}\ \bibnamefont {Lee}},\ }\bibfield  {title} {\enquote
  {\bibinfo {title} {Optimization and control of a small-angle negative ion
  source using an on-line adaptive controller based on the connectionist
  normalized local spline neural network},}\ }\href@noop {} {\bibfield
  {journal} {\bibinfo  {journal} {Nuclear Instruments and Methods in Physics
  Research Section B: Beam Interactions with Materials and Atoms}\ }\textbf
  {\bibinfo {volume} {72}},\ \bibinfo {pages} {271--289} (\bibinfo {year}
  {1992})}\BibitemShut {NoStop}%
\bibitem [{\citenamefont {Mead}\ \emph {et~al.}(1994)\citenamefont {Mead},
  \citenamefont {Brown}, \citenamefont {Jones}, \citenamefont {Bowling},\ and\
  \citenamefont {Barnes}}]{Mead94}%
  \BibitemOpen
  \bibfield  {author} {\bibinfo {author} {\bibfnamefont {W.C.}\ \bibnamefont
  {Mead}}, \bibinfo {author} {\bibfnamefont {S.K.}\ \bibnamefont {Brown}},
  \bibinfo {author} {\bibfnamefont {R.D.}\ \bibnamefont {Jones}}, \bibinfo
  {author} {\bibfnamefont {P.S.}\ \bibnamefont {Bowling}}, \ and\ \bibinfo
  {author} {\bibfnamefont {C.W.}\ \bibnamefont {Barnes}},\ }\bibfield  {title}
  {\enquote {\bibinfo {title} {Adaptive optimization and control using neural
  networks},}\ }\href@noop {} {\bibfield  {journal} {\bibinfo  {journal}
  {Nuclear Instruments and Methods in Physics Research Section A: Accelerators,
  Spectrometers, Detectors and Associated Equipment}\ }\textbf {\bibinfo
  {volume} {352}},\ \bibinfo {pages} {309--315} (\bibinfo {year}
  {1994})}\BibitemShut {NoStop}%
\bibitem [{\citenamefont {Bozoki}\ and\ \citenamefont
  {Friedman}(1994)}]{Bozoki94}%
  \BibitemOpen
  \bibfield  {author} {\bibinfo {author} {\bibfnamefont {E.}~\bibnamefont
  {Bozoki}}\ and\ \bibinfo {author} {\bibfnamefont {A.}~\bibnamefont
  {Friedman}},\ }\bibfield  {title} {\enquote {\bibinfo {title} {Neural network
  technique for orbit correction in accelerators/storage rings},}\ }in\
  \href@noop {} {\emph {\bibinfo {booktitle} {AIP Conference Proceedings}}},\
  Vol.\ \bibinfo {volume} {315}\ (\bibinfo {organization} {AIP},\ \bibinfo
  {year} {1994})\ pp.\ \bibinfo {pages} {103--110}\BibitemShut {NoStop}%
\bibitem [{\citenamefont {Nguyen}\ \emph {et~al.}(1991)\citenamefont {Nguyen},
  \citenamefont {Lee}, \citenamefont {Sass},\ and\ \citenamefont
  {Shoaee}}]{Nguyen91}%
  \BibitemOpen
  \bibfield  {author} {\bibinfo {author} {\bibfnamefont {D.}~\bibnamefont
  {Nguyen}}, \bibinfo {author} {\bibfnamefont {M.}~\bibnamefont {Lee}},
  \bibinfo {author} {\bibfnamefont {R.}~\bibnamefont {Sass}}, \ and\ \bibinfo
  {author} {\bibfnamefont {H.}~\bibnamefont {Shoaee}},\ }\href@noop {} {\emph
  {\bibinfo {title} {Accelerator and feedback control simulation using neural
  networks}}},\ \bibinfo {type} {Tech. Rep.}\ (\bibinfo  {institution}
  {Stanford Linear Accelerator Center, Menlo Park, CA (USA)},\ \bibinfo {year}
  {1991})\BibitemShut {NoStop}%
\bibitem [{\citenamefont {Hitaka}\ \emph {et~al.}(2004)\citenamefont {Hitaka},
  \citenamefont {Shirakata}, \citenamefont {Sato}, \citenamefont {Yokomichi},\
  and\ \citenamefont {Kono}}]{Hitaka04}%
  \BibitemOpen
  \bibfield  {author} {\bibinfo {author} {\bibfnamefont {Y.}~\bibnamefont
  {Hitaka}}, \bibinfo {author} {\bibfnamefont {M.}~\bibnamefont {Shirakata}},
  \bibinfo {author} {\bibfnamefont {H.}~\bibnamefont {Sato}}, \bibinfo {author}
  {\bibfnamefont {M.}~\bibnamefont {Yokomichi}}, \ and\ \bibinfo {author}
  {\bibfnamefont {M.}~\bibnamefont {Kono}},\ }\bibfield  {title} {\enquote
  {\bibinfo {title} {Numerical methods for the orbit control at the kek 12
  gev-ps},}\ }in\ \href@noop {} {\emph {\bibinfo {booktitle} {Proceedings of
  the European Particle Accelerator Conference}}}\ (\bibinfo {year} {2004})\
  pp.\ \bibinfo {pages} {2664--2666}\BibitemShut {NoStop}%
\bibitem [{\citenamefont {Kim}\ \emph {et~al.}(2000)\citenamefont {Kim},
  \citenamefont {Shim}, \citenamefont {Choi}, \citenamefont {Cho},
  \citenamefont {Namkung},\ and\ \citenamefont {Ko}}]{Kim00}%
  \BibitemOpen
  \bibfield  {author} {\bibinfo {author} {\bibfnamefont {K.H.}\ \bibnamefont
  {Kim}}, \bibinfo {author} {\bibfnamefont {K.}~\bibnamefont {Shim}}, \bibinfo
  {author} {\bibfnamefont {J.}~\bibnamefont {Choi}}, \bibinfo {author}
  {\bibfnamefont {M.H.}\ \bibnamefont {Cho}}, \bibinfo {author} {\bibfnamefont
  {W.}~\bibnamefont {Namkung}}, \ and\ \bibinfo {author} {\bibfnamefont {I.S.}\
  \bibnamefont {Ko}},\ }\bibfield  {title} {\enquote {\bibinfo {title}
  {Simulation of the global orbit feedback system for pohang light source},}\
  }in\ \href@noop {} {\emph {\bibinfo {booktitle} {Proceedings of the European
  Particle Accelerator Conference}}}\ (\bibinfo {year} {2000})\ pp.\ \bibinfo
  {pages} {1906--1908}\BibitemShut {NoStop}%
\bibitem [{\citenamefont {Schirmer}\ \emph {et~al.}(2006)\citenamefont
  {Schirmer}, \citenamefont {Hartmann}, \citenamefont {B{\"u}ning},\ and\
  \citenamefont {M{\"u}ller}}]{Schirmer06}%
  \BibitemOpen
  \bibfield  {author} {\bibinfo {author} {\bibfnamefont {D.}~\bibnamefont
  {Schirmer}}, \bibinfo {author} {\bibfnamefont {P.}~\bibnamefont {Hartmann}},
  \bibinfo {author} {\bibfnamefont {T.}~\bibnamefont {B{\"u}ning}}, \ and\
  \bibinfo {author} {\bibfnamefont {D.}~\bibnamefont {M{\"u}ller}},\ }\bibfield
   {title} {\enquote {\bibinfo {title} {Electron transport line optimization
  using neural networks and genetic algorithms},}\ }in\ \href@noop {} {\emph
  {\bibinfo {booktitle} {Proceedings of EPAC}}}\ (\bibinfo {year}
  {2006})\BibitemShut {NoStop}%
\bibitem [{\citenamefont {Jennings}\ \emph {et~al.}(1996)\citenamefont
  {Jennings}, \citenamefont {Mamdani}, \citenamefont {Corera}, \citenamefont
  {Laresgoiti}, \citenamefont {Perriolat}, \citenamefont {Skarek},\ and\
  \citenamefont {Varga}}]{Jennings96}%
  \BibitemOpen
  \bibfield  {author} {\bibinfo {author} {\bibfnamefont {N.R.}\ \bibnamefont
  {Jennings}}, \bibinfo {author} {\bibfnamefont {E.H.}\ \bibnamefont
  {Mamdani}}, \bibinfo {author} {\bibfnamefont {J.M.}\ \bibnamefont {Corera}},
  \bibinfo {author} {\bibfnamefont {I.}~\bibnamefont {Laresgoiti}}, \bibinfo
  {author} {\bibfnamefont {F.}~\bibnamefont {Perriolat}}, \bibinfo {author}
  {\bibfnamefont {P.}~\bibnamefont {Skarek}}, \ and\ \bibinfo {author}
  {\bibfnamefont {L.}~\bibnamefont {Varga}},\ }\bibfield  {title} {\enquote
  {\bibinfo {title} {Using archon to develop real-world dai applications. 1},}\
  }\href@noop {} {\bibfield  {journal} {\bibinfo  {journal} {IEEE expert}\
  }\textbf {\bibinfo {volume} {11}},\ \bibinfo {pages} {64--70} (\bibinfo
  {year} {1996})}\BibitemShut {NoStop}%
\bibitem [{\citenamefont {Perriollat}\ \emph {et~al.}(1996)\citenamefont
  {Perriollat}, \citenamefont {Skarek}, \citenamefont {Varga},\ and\
  \citenamefont {Jennings}}]{Perriollat96}%
  \BibitemOpen
  \bibfield  {author} {\bibinfo {author} {\bibfnamefont {F.}~\bibnamefont
  {Perriollat}}, \bibinfo {author} {\bibfnamefont {P.}~\bibnamefont {Skarek}},
  \bibinfo {author} {\bibfnamefont {L.Z.}\ \bibnamefont {Varga}}, \ and\
  \bibinfo {author} {\bibfnamefont {N.R.}\ \bibnamefont {Jennings}},\
  }\bibfield  {title} {\enquote {\bibinfo {title} {Using archon-3. particle
  acceleration control},}\ }\href@noop {} {\bibfield  {journal} {\bibinfo
  {journal} {IEEE Expert}\ }\textbf {\bibinfo {volume} {11}},\ \bibinfo {pages}
  {80--86} (\bibinfo {year} {1996})}\BibitemShut {NoStop}%
\bibitem [{\citenamefont {Kijima}\ \emph {et~al.}(1992)\citenamefont {Kijima},
  \citenamefont {Yoshida}, \citenamefont {Mizota},\ and\ \citenamefont
  {Suzuki}}]{Kijima92}%
  \BibitemOpen
  \bibfield  {author} {\bibinfo {author} {\bibfnamefont {Y.}~\bibnamefont
  {Kijima}}, \bibinfo {author} {\bibfnamefont {K.}~\bibnamefont {Yoshida}},
  \bibinfo {author} {\bibfnamefont {M.}~\bibnamefont {Mizota}}, \ and\ \bibinfo
  {author} {\bibfnamefont {K.}~\bibnamefont {Suzuki}},\ }\bibfield  {title}
  {\enquote {\bibinfo {title} {A beam diagnostic system for accelerators using
  neural networks},}\ }in\ \href@noop {} {\emph {\bibinfo {booktitle}
  {Proceedings of the European Particle Accelerator Conference}}}\ (\bibinfo
  {year} {1992})\ pp.\ \bibinfo {pages} {1155--1157}\BibitemShut {NoStop}%
\bibitem [{\citenamefont {Stern}\ \emph {et~al.}(1997)\citenamefont {Stern},
  \citenamefont {Klein}, \citenamefont {Luger}, \citenamefont {Kroupa},\ and\
  \citenamefont {Westervelt}}]{Stern97}%
  \BibitemOpen
  \bibfield  {author} {\bibinfo {author} {\bibfnamefont {C.R.}\ \bibnamefont
  {Stern}}, \bibinfo {author} {\bibfnamefont {W.B.}\ \bibnamefont {Klein}},
  \bibinfo {author} {\bibfnamefont {G.F.}\ \bibnamefont {Luger}}, \bibinfo
  {author} {\bibfnamefont {M.}~\bibnamefont {Kroupa}}, \ and\ \bibinfo {author}
  {\bibfnamefont {R.T.}\ \bibnamefont {Westervelt}},\ }\bibfield  {title}
  {\enquote {\bibinfo {title} {Tuning and optimization at brookhaven and
  argonne: Results of recent experiments using a portable intelligent control
  system},}\ }in\ \href@noop {} {\emph {\bibinfo {booktitle} {APS Meeting
  Abstracts}}}\ (\bibinfo {year} {1997})\BibitemShut {NoStop}%
\bibitem [{\citenamefont {Klein}\ \emph
  {et~al.}(1997{\natexlab{a}})\citenamefont {Klein}, \citenamefont {Stern},
  \citenamefont {Luger},\ and\ \citenamefont {Olsson}}]{Klein97}%
  \BibitemOpen
  \bibfield  {author} {\bibinfo {author} {\bibfnamefont {W.B.}\ \bibnamefont
  {Klein}}, \bibinfo {author} {\bibfnamefont {C.R.}\ \bibnamefont {Stern}},
  \bibinfo {author} {\bibfnamefont {G.F.}\ \bibnamefont {Luger}}, \ and\
  \bibinfo {author} {\bibfnamefont {E.T.}\ \bibnamefont {Olsson}},\ }\bibfield
  {title} {\enquote {\bibinfo {title} {An intelligent control architecture for
  accelerator beamline tuning},}\ }in\ \href@noop {} {\emph {\bibinfo
  {booktitle} {Proceedings of the fourteenth national conference on artificial
  intelligence and ninth conference on Innovative applications of artificial
  intelligence}}}\ (\bibinfo {organization} {AAAI Press},\ \bibinfo {year}
  {1997})\ pp.\ \bibinfo {pages} {1019--1025}\BibitemShut {NoStop}%
\bibitem [{\citenamefont {Klein}\ \emph
  {et~al.}(1997{\natexlab{b}})\citenamefont {Klein}, \citenamefont {Stern},
  \citenamefont {Luger},\ and\ \citenamefont {Olsson}}]{Klein97-2}%
  \BibitemOpen
  \bibfield  {author} {\bibinfo {author} {\bibfnamefont {W.}~\bibnamefont
  {Klein}}, \bibinfo {author} {\bibfnamefont {C.}~\bibnamefont {Stern}},
  \bibinfo {author} {\bibfnamefont {G.}~\bibnamefont {Luger}}, \ and\ \bibinfo
  {author} {\bibfnamefont {E.}~\bibnamefont {Olsson}},\ }\bibfield  {title}
  {\enquote {\bibinfo {title} {Designing a portable architecture for
  intelligent particle accelerator control},}\ }in\ \href@noop {} {\emph
  {\bibinfo {booktitle} {Particle Accelerator Conference, 1997. Proceedings of
  the 1997}}},\ Vol.~\bibinfo {volume} {2}\ (\bibinfo {organization} {IEEE},\
  \bibinfo {year} {1997})\ pp.\ \bibinfo {pages} {2422--2424}\BibitemShut
  {NoStop}%
\bibitem [{\citenamefont {Klein}\ \emph {et~al.}(1999)\citenamefont {Klein},
  \citenamefont {Westervelt},\ and\ \citenamefont {Luger}}]{Klein99}%
  \BibitemOpen
  \bibfield  {author} {\bibinfo {author} {\bibfnamefont {W.}~\bibnamefont
  {Klein}}, \bibinfo {author} {\bibfnamefont {R.}~\bibnamefont {Westervelt}}, \
  and\ \bibinfo {author} {\bibfnamefont {G.}~\bibnamefont {Luger}},\ }\bibfield
   {title} {\enquote {\bibinfo {title} {Developing a general purpose
  intelligent control system for particle accelerators},}\ }\href@noop {}
  {\bibfield  {journal} {\bibinfo  {journal} {Journal of Intelligent \& Fuzzy
  Systems}\ }\textbf {\bibinfo {volume} {7}},\ \bibinfo {pages} {1--12}
  (\bibinfo {year} {1999})}\BibitemShut {NoStop}%
\bibitem [{\citenamefont {Wielgosz}\ \emph {et~al.}(2017)\citenamefont
  {Wielgosz}, \citenamefont {Skoczeń},\ and\ \citenamefont
  {Mertik}}]{Wielgosz:2016xhl}%
  \BibitemOpen
  \bibfield  {author} {\bibinfo {author} {\bibfnamefont {M.}~\bibnamefont
  {Wielgosz}}, \bibinfo {author} {\bibfnamefont {A.}~\bibnamefont {Skoczeń}},
  \ and\ \bibinfo {author} {\bibfnamefont {M.}~\bibnamefont {Mertik}},\
  }\bibfield  {title} {\enquote {\bibinfo {title} {{Using LSTM recurrent neural
  networks for monitoring the LHC superconducting magnets}},}\ }\href {\doibase
  10.1016/j.nima.2017.06.020} {\bibfield  {journal} {\bibinfo  {journal} {Nucl.
  Instrum. Meth.}\ }\textbf {\bibinfo {volume} {A867}},\ \bibinfo {pages}
  {40--50} (\bibinfo {year} {2017})},\ \Eprint
  {http://arxiv.org/abs/1611.06241} {arXiv:1611.06241 [physics.ins-det]}
  \BibitemShut {NoStop}%
\bibitem [{\citenamefont {Nawaz}\ \emph {et~al.}(2016)\citenamefont {Nawaz},
  \citenamefont {Pfeiffer}, \citenamefont {Lichtenberg},\ and\ \citenamefont
  {Schlarb}}]{NawazDESY2016}%
  \BibitemOpen
  \bibfield  {author} {\bibinfo {author} {\bibfnamefont {A.~S.}\ \bibnamefont
  {Nawaz}}, \bibinfo {author} {\bibfnamefont {S.}~\bibnamefont {Pfeiffer}},
  \bibinfo {author} {\bibfnamefont {G.}~\bibnamefont {Lichtenberg}}, \ and\
  \bibinfo {author} {\bibfnamefont {H.}~\bibnamefont {Schlarb}},\ }\bibfield
  {title} {\enquote {\bibinfo {title} {Self-organzied critical control for the
  european xfel using black box parameter identification for the quench
  detection system},}\ }in\ \href {\doibase 10.1109/SYSTOL.2016.7739750} {\emph
  {\bibinfo {booktitle} {2016 3rd Conference on Control and Fault-Tolerant
  Systems (SysTol)}}}\ (\bibinfo {year} {2016})\ pp.\ \bibinfo {pages}
  {196--201}\BibitemShut {NoStop}%
\bibitem [{\citenamefont {Fol}\ \emph {et~al.}(2017)\citenamefont {Fol},
  \citenamefont {Tomas~Garcia},\ and\ \citenamefont {Henning}}]{Fol:2309558}%
  \BibitemOpen
  \bibfield  {author} {\bibinfo {author} {\bibfnamefont {E.}~\bibnamefont
  {Fol}}, \bibinfo {author} {\bibfnamefont {R.}~\bibnamefont {Tomas~Garcia}}, \
  and\ \bibinfo {author} {\bibfnamefont {P.}~\bibnamefont {Henning}},\ }\\\href
  {http://cds.cern.ch/record/2309558} {\enquote {\bibinfo {title} {{Evaluation
  of Machine Learning Methods for LHC Optics Measurements and Corrections
  Software}},}\ } (\bibinfo {year} {2017}),\ \bibinfo {note} {presented 23 Oct
  2017}\BibitemShut {NoStop}%
\bibitem [{\citenamefont {Valentino}\ \emph {et~al.}(2017)\citenamefont
  {Valentino}, \citenamefont {Bruce}, \citenamefont {Redaelli}, \citenamefont
  {Rossi}, \citenamefont {Theodoropoulos},\ and\ \citenamefont
  {Jaster-Merz}}]{Valentino:2017hlm}%
  \BibitemOpen
  \bibfield  {author} {\bibinfo {author} {\bibfnamefont {G.}~\bibnamefont
  {Valentino}}, \bibinfo {author} {\bibfnamefont {R.}~\bibnamefont {Bruce}},
  \bibinfo {author} {\bibfnamefont {S.}~\bibnamefont {Redaelli}}, \bibinfo
  {author} {\bibfnamefont {R.}~\bibnamefont {Rossi}}, \bibinfo {author}
  {\bibfnamefont {P.}~\bibnamefont {Theodoropoulos}}, \ and\ \bibinfo {author}
  {\bibfnamefont {S.}~\bibnamefont {Jaster-Merz}},\ }\bibfield  {title}
  {\enquote {\bibinfo {title} {{Anomaly Detection for Beam Loss Maps in the
  Large Hadron Collider}},}\ }\bibfield  {booktitle} {\emph {\bibinfo
  {booktitle} {{Proceedings, 8th International Particle Accelerator Conference
  (IPAC 2017): Copenhagen, Denmark, May 14-19, 2017}}},\ }\href {\doibase
  10.1088/1742-6596/874/1/012002, 10.18429/JACoW-IPAC2017-MOPAB010} {\bibfield
  {journal} {\bibinfo  {journal} {J. Phys. Conf. Ser.}\ }\textbf {\bibinfo
  {volume} {874}},\ \bibinfo {pages} {012002} (\bibinfo {year} {2017})},\
  \bibinfo {note} {[,MOPAB010(2017)]}\BibitemShut {NoStop}%
\bibitem [{\citenamefont {King}\ \emph {et~al.}(2017)\citenamefont {King},
  \citenamefont {Pogorelov}, \citenamefont {Amyx}, \citenamefont {Borland},\
  and\ \citenamefont {Soliday}}]{King:2017vfo}%
  \BibitemOpen
  \bibfield  {author} {\bibinfo {author} {\bibfnamefont {J.~R.}\ \bibnamefont
  {King}}, \bibinfo {author} {\bibfnamefont {I.~V.}\ \bibnamefont {Pogorelov}},
  \bibinfo {author} {\bibfnamefont {K.~M.}\ \bibnamefont {Amyx}}, \bibinfo
  {author} {\bibfnamefont {M.}~\bibnamefont {Borland}}, \ and\ \bibinfo
  {author} {\bibfnamefont {R.}~\bibnamefont {Soliday}},\ }\bibfield  {title}
  {\enquote {\bibinfo {title} {{GPU acceleration and performance of the
  particle-beam-dynamics code Elegant}},}\ }\href@noop {} {\  (\bibinfo {year}
  {2017})},\ \Eprint {http://arxiv.org/abs/1710.07350} {arXiv:1710.07350
  [physics.comp-ph]} \BibitemShut {NoStop}%
\bibitem [{\citenamefont {Pang}\ and\ \citenamefont
  {Rybarcyk}(2014)}]{Pang:2014cda}%
  \BibitemOpen
  \bibfield  {author} {\bibinfo {author} {\bibfnamefont {X.}~\bibnamefont
  {Pang}}\ and\ \bibinfo {author} {\bibfnamefont {L.}~\bibnamefont
  {Rybarcyk}},\ }\bibfield  {title} {\enquote {\bibinfo {title} {{GPU
  accelerated online multi-particle beam dynamics simulator for ion linear
  particle accelerators}},}\ }\href {\doibase 10.1016/j.cpc.2013.10.033}
  {\bibfield  {journal} {\bibinfo  {journal} {Comput. Phys. Commun.}\ }\textbf
  {\bibinfo {volume} {185}},\ \bibinfo {pages} {744--753} (\bibinfo {year}
  {2014})}\BibitemShut {NoStop}%
\bibitem [{\citenamefont {Pang}(2015)}]{Pang:2015iul}%
  \BibitemOpen
  \bibfield  {author} {\bibinfo {author} {\bibfnamefont {X.}~\bibnamefont
  {Pang}},\ }\bibfield  {title} {\enquote {\bibinfo {title} {{Advances in
  Proton Linac Online Modeling}},}\ }in\ \href
  {http://accelconf.web.cern.ch/AccelConf/IPAC2015/papers/wexc2.pdf} {\emph
  {\bibinfo {booktitle} {{Proceedings, 6th International Particle Accelerator
  Conference (IPAC 2015)}}}}\ (\bibinfo
  {year} {2015})\ p.\ \bibinfo {pages} {WEXC2}\BibitemShut {NoStop}%
\bibitem [{\citenamefont {Edelen}\ \emph
  {et~al.}(2018{\natexlab{a}})\citenamefont {Edelen}, \citenamefont {Edelen},
  \citenamefont {Bowring}, \citenamefont {Chase}, \citenamefont {Edstrom},
  \citenamefont {Steimel}, \citenamefont {van~der Slot},\ and\ \citenamefont
  {Biedron}}]{Edel18_2}%
  \BibitemOpen
  \bibfield  {author} {\bibinfo {author} {\bibfnamefont {A.~L.}\ \bibnamefont
  {Edelen}}, \bibinfo {author} {\bibfnamefont {J.P.}\ \bibnamefont {Edelen}},
  \bibinfo {author} {\bibfnamefont {D.}~\bibnamefont {Bowring}}, \bibinfo
  {author} {\bibfnamefont {B.E.}\ \bibnamefont {Chase}}, \bibinfo {author}
  {\bibfnamefont {D.R.}\ \bibnamefont {Edstrom}}, \bibinfo {author}
  {\bibfnamefont {J.}~\bibnamefont {Steimel}}, \bibinfo {author} {\bibfnamefont
  {P.J.M.}\ \bibnamefont {van~der Slot}}, \ and\ \bibinfo {author}
  {\bibfnamefont {S.G.}\ \bibnamefont {Biedron}},\ }\bibfield  {title}
  {\enquote {\bibinfo {title} {{Recent Applications of Neural Network-Based
  Approaches to the Modeling and Control of Particle Accelerators}},}\ }in\
  \href@noop {} {\emph {\bibinfo {booktitle} {{9th International Particle
  Accelerator Conference (IPAC 2018): Vancouver, BC, Canada, April 29-May 4,
  2018}}}}\ (\bibinfo {year} {2018})\ p.\ \bibinfo {pages}
  {THYGBE2}\BibitemShut {NoStop}%
\bibitem [{\citenamefont {Edelen}\ \emph
  {et~al.}(2017{\natexlab{a}})\citenamefont {Edelen}, \citenamefont {Biedron},
  \citenamefont {Edelen},\ and\ \citenamefont {Milton}}]{Edelen:2017ewy}%
  \BibitemOpen
  \bibfield  {author} {\bibinfo {author} {\bibfnamefont {Auralee}\ \bibnamefont
  {Edelen}}, \bibinfo {author} {\bibfnamefont {Sandra}\ \bibnamefont
  {Biedron}}, \bibinfo {author} {\bibfnamefont {Jonathan}\ \bibnamefont
  {Edelen}}, \ and\ \bibinfo {author} {\bibfnamefont {Stephen}\ \bibnamefont
  {Milton}},\ }\bibfield  {title} {\enquote {\bibinfo {title} {{First Steps
  Toward Incorporating Image Based Diagnostics into Particle Accelerator
  Control Systems Using Convolutional Neural Networks}},}\ }in\ \href {\doibase
  10.18429/JACoW-NAPAC2016-TUPOA51} {\emph {\bibinfo {booktitle} {{Proceedings,
  2nd North American Particle Accelerator Conference (NAPAC2016)}}}}\ (\bibinfo {year} {2017})\ p.\ \bibinfo
  {pages} {TUPOA51},\ \Eprint {http://arxiv.org/abs/1612.05662}
  {arXiv:1612.05662 [physics.acc-ph]} \BibitemShut {NoStop}%
\bibitem [{\citenamefont {Adelmann}(2015)}]{Adelmann:2015wva}%
  \BibitemOpen
  \bibfield  {author} {\bibinfo {author} {\bibfnamefont {A.}~\bibnamefont
  {Adelmann}},\ }\bibfield  {title} {\enquote {\bibinfo {title} {{On
  Uncertainty Quantification in Particle Accelerators Modelling}},}\
  }\href@noop {} {\  (\bibinfo {year} {2015})},\ \Eprint
  {http://arxiv.org/abs/1509.08130} {arXiv:1509.08130 [physics.acc-ph]}
  \BibitemShut {NoStop}%
\bibitem [{\citenamefont {Edelen}\ \emph
  {et~al.}(2017{\natexlab{b}})\citenamefont {Edelen}, \citenamefont {Biedron},
  \citenamefont {Edelen}, \citenamefont {Milton},\ and\ \citenamefont {{van der
  Slot}}}]{auraleeFEL2017}%
  \BibitemOpen
  \bibfield  {author} {\bibinfo {author} {\bibfnamefont {A.L.}\ \bibnamefont
  {Edelen}}, \bibinfo {author} {\bibfnamefont {S.G.}\ \bibnamefont {Biedron}},
  \bibinfo {author} {\bibfnamefont {J.P.}\ \bibnamefont {Edelen}}, \bibinfo
  {author} {\bibfnamefont {S.V.}\ \bibnamefont {Milton}}, \ and\ \bibinfo
  {author} {\bibfnamefont {P.J.M.}\ \bibnamefont {{van der Slot}}},\ }\enquote
  {\bibinfo {title} {Using a neural network control policy for rapid switching
  between beam parameters in an fel},}\ in\ \href@noop {} {\emph {\bibinfo
  {booktitle} {Proceedings of the 38th International Free Electron Laser
  Conference}}}\ (\bibinfo {year} {2017})\ pp.\ \bibinfo {pages}
  {406--409}\BibitemShut {NoStop}%
\bibitem [{\citenamefont {{Y. Wang, M. Borland, H. Shang, R. Soliday, A.
  Xiao}}()}]{pelegant}%
  \BibitemOpen
  \bibfield  {author} {\bibinfo {author} {\bibnamefont {{Y. Wang, M. Borland,
  H. Shang, R. Soliday, A. Xiao}}},\ }\bibfield  {title} {\enquote {\bibinfo
  {title} {{Recent Progress on Parallel ELEGANT}},}\ }in\ \href@noop {} {\emph
  {\bibinfo {booktitle} {{Proceeding ICAP 2009: San Francisco, CA
  (2009)}}}}\BibitemShut {NoStop}%
\bibitem [{\citenamefont {{Y. Ineichen, A. Adelmann, C. Bekas, A. Curioni, P.
  Arbenz}}(2013)}]{opal_parallel}%
  \BibitemOpen
  \bibfield  {author} {\bibinfo {author} {\bibnamefont {{Y. Ineichen, A.
  Adelmann, C. Bekas, A. Curioni, P. Arbenz}}},\ }\bibfield  {title} {\enquote
  {\bibinfo {title} {{A fast and scalable low dimensional solver for charged
  particle dynamics in large particle accelerators}},}\ }\href {\doibase
  10.1007/s00450-012-0216-2} {\bibfield  {journal} {\bibinfo  {journal}
  {{Comput. Sci. Res. Dev.,}}\ }\textbf {\bibinfo {volume} {28}},\ \bibinfo
  {pages} {185--192} (\bibinfo {year} {2013})}\BibitemShut {NoStop}%
\bibitem [{\citenamefont {Huang}\ and\ \citenamefont
  {Safranek}(2015)}]{Huang:2015wka}%
  \BibitemOpen
  \bibfield  {author} {\bibinfo {author} {\bibfnamefont {X.}~\bibnamefont
  {Huang}}\ and\ \bibinfo {author} {\bibfnamefont {J.}~\bibnamefont
  {Safranek}},\ }\bibfield  {title} {\enquote {\bibinfo {title} {{Online
  optimization of storage ring nonlinear beam dynamics}},}\ }\href {\doibase
  10.1103/PhysRevSTAB.18.084001} {\bibfield  {journal} {\bibinfo  {journal}
  {Phys. Rev. ST Accel. Beams}\ }\textbf {\bibinfo {volume} {18}},\ \bibinfo
  {pages} {084001} (\bibinfo {year} {2015})},\ \Eprint
  {http://arxiv.org/abs/1502.07799} {arXiv:1502.07799 [physics.acc-ph]}
  \BibitemShut {NoStop}%
\bibitem [{\citenamefont {Scheinker}\ and\ \citenamefont
  {Gessner}(2015)}]{PhysRevSTAB.18.102801}%
  \BibitemOpen
  \bibfield  {author} {\bibinfo {author} {\bibfnamefont {A.}~\bibnamefont
  {Scheinker}}\ and\ \bibinfo {author} {\bibfnamefont {S.}~\bibnamefont
  {Gessner}},\ }\bibfield  {title} {\enquote {\bibinfo {title} {Adaptive method
  for electron bunch profile prediction},}\ }\href {\doibase
  10.1103/PhysRevSTAB.18.102801} {\bibfield  {journal} {\bibinfo  {journal}
  {Phys. Rev. ST Accel. Beams}\ }\textbf {\bibinfo {volume} {18}},\ \bibinfo
  {pages} {102801} (\bibinfo {year} {2015})}\BibitemShut {NoStop}%
\bibitem [{\citenamefont {Sanchez-Gonzalez}\ \emph {et~al.}(2017)\citenamefont
  {Sanchez-Gonzalez}, \citenamefont {Micaelli}, \citenamefont {Olivier},
  \citenamefont {Barillot}, \citenamefont {Ilchen}, \citenamefont {Lutman},
  \citenamefont {Marinelli}, \citenamefont {Maxwell}, \citenamefont {Achner},
  \citenamefont {Ag{\aa}ker} \emph {et~al.}}]{Sanchez-Gonzalez2017}%
  \BibitemOpen
  \bibfield  {author} {\bibinfo {author} {\bibfnamefont {A.}~\bibnamefont
  {Sanchez-Gonzalez}}, \bibinfo {author} {\bibfnamefont {P.}~\bibnamefont
  {Micaelli}}, \bibinfo {author} {\bibfnamefont {C.}~\bibnamefont {Olivier}},
  \bibinfo {author} {\bibfnamefont {T.~R.}\ \bibnamefont {Barillot}}, \bibinfo
  {author} {\bibfnamefont {M.}~\bibnamefont {Ilchen}}, \bibinfo {author}
  {\bibfnamefont {A.~A.}\ \bibnamefont {Lutman}}, \bibinfo {author}
  {\bibfnamefont {A.}~\bibnamefont {Marinelli}}, \bibinfo {author}
  {\bibfnamefont {T.}~\bibnamefont {Maxwell}}, \bibinfo {author} {\bibfnamefont
  {A.}~\bibnamefont {Achner}}, \bibinfo {author} {\bibfnamefont
  {M.}~\bibnamefont {Ag{\aa}ker}},  \emph {et~al.},\ }\bibfield  {title}
  {\enquote {\bibinfo {title} {Accurate prediction of x-ray pulse properties
  from a free-electron laser using machine learning},}\ }\href
  {http://dx.doi.org/10.1038/ncomms15461} {\bibfield  {journal} {\bibinfo
  {journal} {Nature Communications}\ }\textbf {\bibinfo {volume} {8}},\
  \bibinfo {pages} {15461 EP --} (\bibinfo {year} {2017})}\BibitemShut
  {NoStop}%
\bibitem [{\citenamefont {Edelen}\ \emph
  {et~al.}(2018{\natexlab{b}})\citenamefont {Edelen}, \citenamefont {Edelen},
  \citenamefont {Edstrom}, \citenamefont {Halavanau}, \citenamefont {Piot},\
  and\ \citenamefont {Biedron}}]{Edel18_1}%
  \BibitemOpen
  \bibfield  {author} {\bibinfo {author} {\bibfnamefont {A.~L.}\ \bibnamefont
  {Edelen}}, \bibinfo {author} {\bibfnamefont {J.P.}\ \bibnamefont {Edelen}},
  \bibinfo {author} {\bibfnamefont {D.R.}\ \bibnamefont {Edstrom}}, \bibinfo
  {author} {\bibfnamefont {A.}~\bibnamefont {Halavanau}}, \bibinfo {author}
  {\bibfnamefont {P.}~\bibnamefont {Piot}}, \ and\ \bibinfo {author}
  {\bibfnamefont {S.G.}\ \bibnamefont {Biedron}},\ }\bibfield  {title}
  {\enquote {\bibinfo {title} {{Neural Network Virtual Diagnostic for the FAST
  Low Energy Beamline}},}\ }in\ \href@noop {} {\emph {\bibinfo {booktitle}
  {{9th International Particle Accelerator Conference (IPAC 2018): Vancouver,
  BC, Canada, April 29-May 4, 2018}}}}\ (\bibinfo {year} {2018})\ p.\ \bibinfo
  {pages} {WEPAF040}\BibitemShut {NoStop}%
\bibitem [{\citenamefont {Emma}\ \emph {et~al.}(2018)\citenamefont {Emma},
  \citenamefont {Edelen}, \citenamefont {White},\ and\ \citenamefont
  {Hogan}}]{AACVD}%
  \BibitemOpen
  \bibfield  {author} {\bibinfo {author} {\bibfnamefont {C.}~\bibnamefont
  {Emma}}, \bibinfo {author} {\bibfnamefont {A.L.}\ \bibnamefont {Edelen}},
  \bibinfo {author} {\bibfnamefont {G.}~\bibnamefont {White}}, \ and\ \bibinfo
  {author} {\bibfnamefont {M.}~\bibnamefont {Hogan}},\ }\bibfield  {title}
  {\enquote {\bibinfo {title} {{Virtual diagnostic for phase space prediction
  at FACET-II}},}\ }in\ \href@noop {} {\emph {\bibinfo {booktitle} {{Advanced
  Accelerator Concepts Workshop: Breckenridge, CO, USA, August 12 - 17,
  2018}}}}\ (\bibinfo {year} {2018})\BibitemShut {NoStop}%
\bibitem [{\citenamefont {{C. Emma, A. L. Edelen, G. White, M. Hogan, B.
  O'Shea, V. Yakimenko}}(Oct. 2018)}]{prab_tcav}%
  \BibitemOpen
  \bibfield  {author} {\bibinfo {author} {\bibnamefont {{C. Emma, A. L. Edelen,
  G. White, M. Hogan, B. O'Shea, V. Yakimenko}}},\ }\bibfield  {title}
  {\enquote {\bibinfo {title} {{Machine Learning-based Longitudinal Phase Space
  Prediction of Particle Accelerators}},}\ }\href@noop {} {\bibfield  {journal}
  {\bibinfo  {journal} {{accepted in Physical Review Accelerators and Beams,}}\
  } (\bibinfo {year} {Oct. 2018})}\BibitemShut {NoStop}%
\bibitem [{\citenamefont {Wu}\ \emph {et~al.}(2018)\citenamefont {Wu},
  \citenamefont {Huang}, \citenamefont {Raubenheimer},\ and\ \citenamefont
  {Scheinker}}]{Wu:2018tdh}%
  \BibitemOpen
  \bibfield  {author} {\bibinfo {author} {\bibfnamefont {J.}~\bibnamefont
  {Wu}}, \bibinfo {author} {\bibfnamefont {X.}~\bibnamefont {Huang}}, \bibinfo
  {author} {\bibfnamefont {T.}~\bibnamefont {Raubenheimer}}, \ and\ \bibinfo
  {author} {\bibfnamefont {A.}~\bibnamefont {Scheinker}},\ }\bibfield  {title}
  {\enquote {\bibinfo {title} {{Recent On-Line Taper Optimization on LCLS}},}\
  }in\ \href {\doibase 10.18429/JACoW-FEL2017-TUB04} {\emph {\bibinfo
  {booktitle} {{Proceedings, 38th International Free Electron Laser Conference,
  FEL2017}}}}\ (\bibinfo {year} {2018})\ p.\ \bibinfo {pages}
  {TUB04}\BibitemShut {NoStop}%
\bibitem [{\citenamefont {McIntire}\ \emph {et~al.}(2016)\citenamefont
  {McIntire}, \citenamefont {Cope}, \citenamefont {Ermon},\ and\ \citenamefont
  {Ratner}}]{McIntire:2016fnl}%
  \BibitemOpen
  \bibfield  {author} {\bibinfo {author} {\bibfnamefont {M.}~\bibnamefont
  {McIntire}}, \bibinfo {author} {\bibfnamefont {T.}~\bibnamefont {Cope}},
  \bibinfo {author} {\bibfnamefont {S.}~\bibnamefont {Ermon}}, \ and\ \bibinfo
  {author} {\bibfnamefont {D.}~\bibnamefont {Ratner}},\ }\bibfield  {title}
  {\enquote {\bibinfo {title} {{Bayesian Optimization of FEL Performance at
  LCLS}},}\ }in\ \href {\doibase 10.18429/JACoW-IPAC2016-WEPOW055} {\emph
  {\bibinfo {booktitle} {{Proceedings, 7th International Particle Accelerator
  Conference (IPAC 2016)}}}}\ (\bibinfo {year}
  {2016})\ p.\ \bibinfo {pages} {WEPOW055}\BibitemShut {NoStop}%
\bibitem [{\citenamefont {Berkenkamp}\ \emph
  {et~al.}(2016{\natexlab{a}})\citenamefont {Berkenkamp}, \citenamefont
  {Schoellig},\ and\ \citenamefont {Krause}}]{Krause1}%
  \BibitemOpen
  \bibfield  {author} {\bibinfo {author} {\bibfnamefont {F.}~\bibnamefont
  {Berkenkamp}}, \bibinfo {author} {\bibfnamefont {A.~P.}\ \bibnamefont
  {Schoellig}}, \ and\ \bibinfo {author} {\bibfnamefont {A.}~\bibnamefont
  {Krause}},\ }\bibfield  {title} {\enquote {\bibinfo {title} {{Safe Controller
  Optimization for Quadrotors with Gaussian Processes}},} }
  {\bibfield  {journal}
  {\bibinfo  {journal} {Proc. of the IEEE Int. Conf. on Robotics
  and Automation (ICRA)}\ } (\bibinfo {year} {2016}{\natexlab{a}}),\
  https://doi.org/10.1088/1748-0221/11/09/P09001}\BibitemShut {NoStop}%
\bibitem [{\citenamefont {Berkenkamp}\ \emph
  {et~al.}(2016{\natexlab{b}})\citenamefont {Berkenkamp}, \citenamefont
  {Krause},\ and\ \citenamefont {Schoellig}}]{Krause2}%
  \BibitemOpen
  \bibfield  {author} {\bibinfo {author} {\bibfnamefont {F.}~\bibnamefont
  {Berkenkamp}}, \bibinfo {author} {\bibfnamefont {A.}~\bibnamefont {Krause}},
  \ and\ \bibinfo {author} {\bibfnamefont {A.~P.}\ \bibnamefont {Schoellig}},\
  }\bibfield  {title} {\enquote {\bibinfo {title} {{Bayesian Optimization with
  Safety Constraints: Safe and Automatic Parameter Tuning in Robotics}},}\
  }\href {\doibase arXiv:1602.04450} {\bibfield  {journal} {\bibinfo  {journal}
  {arXiv}\ } (\bibinfo {year} {2016}{\natexlab{b}}),\
  arXiv:1602.04450}\BibitemShut {NoStop}%
\bibitem [{\citenamefont {Edelen}\ \emph {et~al.}(2015)\citenamefont {Edelen}
  \emph {et~al.}}]{Edelen:2015lbj}%
  \BibitemOpen
  \bibfield  {author} {\bibinfo {author} {\bibfnamefont {A.~L.}\ \bibnamefont
  {Edelen}} \emph {et~al.},\ }\bibfield  {title} {\enquote {\bibinfo {title}
  {{Initial experimental results of a machine learning-based temperature
  control system for an RF gun}},}\ }in\ \href
  {http://lss.fnal.gov/archive/2015/conf/fermilab-conf-15-182-ad.pdf} {\emph
  {\bibinfo {booktitle} {{Proceedings, 6th International Particle Accelerator
  Conference (IPAC 2015)}}}}\ (\bibinfo
  {year} {2015})\ p.\ \bibinfo {pages} {MOPWI028},\ \Eprint
  {http://arxiv.org/abs/1511.01883} {arXiv:1511.01883 [physics.acc-ph]}
  \BibitemShut {NoStop}%
\bibitem [{\citenamefont {Edelen}\ \emph
  {et~al.}(2016{\natexlab{b}})\citenamefont {Edelen}, \citenamefont {Biedron},
  \citenamefont {Bowring}, \citenamefont {Chase}, \citenamefont {Edelen},
  \citenamefont {Milton},\ and\ \citenamefont {Steimel}}]{Edelen:2016jgh}%
  \BibitemOpen
  \bibfield  {author} {\bibinfo {author} {\bibfnamefont {A.}~\bibnamefont
  {Edelen}}, \bibinfo {author} {\bibfnamefont {S.}~\bibnamefont {Biedron}},
  \bibinfo {author} {\bibfnamefont {D.}~\bibnamefont {Bowring}}, \bibinfo
  {author} {\bibfnamefont {B.}~\bibnamefont {Chase}}, \bibinfo {author}
  {\bibfnamefont {J.}~\bibnamefont {Edelen}}, \bibinfo {author} {\bibfnamefont
  {S.}~\bibnamefont {Milton}}, \ and\ \bibinfo {author} {\bibfnamefont
  {J.}~\bibnamefont {Steimel}},\ }\bibfield  {title} {\enquote {\bibinfo
  {title} {{Neural Network Model Of The PXIE RFQ Cooling System and Resonant
  Frequency Response}},}\ }in\ \href {\doibase
  10.18429/JACoW-IPAC2016-THPOY020} {\emph {\bibinfo {booktitle} {{Proceedings,
  7th International Particle Accelerator Conference (IPAC 2016)}}}}\ (\bibinfo {year} {2016})\ p.\ \bibinfo {pages}
  {THPOY020},\ \Eprint {http://arxiv.org/abs/1612.07237} {arXiv:1612.07237
  [physics.acc-ph]} \BibitemShut {NoStop}%
\bibitem [{\citenamefont {Kong}\ \emph {et~al.}(2016)\citenamefont {Kong},
  \citenamefont {Hur}, \citenamefont {Lee}, \citenamefont {Park}, \citenamefont
  {Park},\ and\ \citenamefont {Yang}}]{KONG201655}%
  \BibitemOpen
  \bibfield  {author} {\bibinfo {author} {\bibfnamefont {Y.B.}\ \bibnamefont
  {Kong}}, \bibinfo {author} {\bibfnamefont {M.G.}\ \bibnamefont {Hur}},
  \bibinfo {author} {\bibfnamefont {E.J.}\ \bibnamefont {Lee}}, \bibinfo
  {author} {\bibfnamefont {J.H.}\ \bibnamefont {Park}}, \bibinfo {author}
  {\bibfnamefont {Y.D.}\ \bibnamefont {Park}}, \ and\ \bibinfo {author}
  {\bibfnamefont {S.D.}\ \bibnamefont {Yang}},\ }\bibfield  {title} {\enquote
  {\bibinfo {title} {Predictive ion source control using artificial neural
  network for rft-30 cyclotron},}\ }\href {\doibase
  https://doi.org/10.1016/j.nima.2015.09.095} {\bibfield  {journal} {\bibinfo
  {journal} {Nucl. Instrum. Methods Phys. Res. A}\ }\textbf
  {\bibinfo {volume} {806}},\ \bibinfo {pages} {55 -- 60} (\bibinfo {year}
  {2016})}\BibitemShut {NoStop}%
\bibitem [{\citenamefont {Mohayai}\ \emph {et~al.}(2018)\citenamefont
  {Mohayai}, \citenamefont {Snopok},\ and\ \citenamefont
  {Neuffer}}]{Mohayai:2018grg}%
  \BibitemOpen
  \bibfield  {author} {\bibinfo {author} {\bibfnamefont {T.A.}\ \bibnamefont
  {Mohayai}}, \bibinfo {author} {\bibfnamefont {P.}~\bibnamefont {Snopok}}, \
  and\ \bibinfo {author} {\bibfnamefont {D.}~\bibnamefont {Neuffer}} (\bibinfo
  {collaboration} {MICE}),\ }\bibfield  {title} {\enquote {\bibinfo {title} {{A
  Non-Parametric Density Estimation Approach to Measuring Beam Cooling in
  MICE}},}\ }in\ \href@noop {} {\emph {\bibinfo {booktitle} {{9th International
  Particle Accelerator Conference (IPAC 2018) Vancouver, BC Canada, April
  29-May 4, 2018}}}}\ (\bibinfo {year} {2018})\ p.\ \bibinfo {pages}
  {TUPML063},\ \Eprint {http://arxiv.org/abs/1806.01834} {arXiv:1806.01834
  [physics.acc-ph]} \BibitemShut {NoStop}%
\bibitem [{\citenamefont {Mohayai}(2018)}]{Mohayai:2018rxn}%
  \BibitemOpen
  \bibfield  {author} {\bibinfo {author} {\bibfnamefont {T.A.}\ \bibnamefont
  {Mohayai}} (\bibinfo {collaboration} {MICE}),\ }\bibfield  {title} {\enquote
  {\bibinfo {title} {{First Demonstration of Ionization Cooling in MICE}},}\
  }in\ \href@noop {} {\emph {\bibinfo {booktitle} {{9th International Particle
  Accelerator Conference (IPAC 2018) Vancouver, BC Canada, April 29-May 4,
  2018}}}}\ (\bibinfo {year} {2018})\ p.\ \bibinfo {pages} {FRXGBE3},\ \Eprint
  {http://arxiv.org/abs/1806.01807} {arXiv:1806.01807 [physics.acc-ph]}
  \BibitemShut {NoStop}%
\bibitem [{\citenamefont {Li}\ \emph {et~al.}(2018)\citenamefont {Li},
  \citenamefont {Cropp}, \citenamefont {Kabra}, \citenamefont {Lane},
  \citenamefont {Wetzstein}, \citenamefont {Musumeci},\ and\ \citenamefont
  {Ratner}}]{sli}%
  \BibitemOpen
  \bibfield  {author} {\bibinfo {author} {\bibfnamefont {S.}~\bibnamefont
  {Li}}, \bibinfo {author} {\bibfnamefont {F.}~\bibnamefont {Cropp}}, \bibinfo
  {author} {\bibfnamefont {K.}~\bibnamefont {Kabra}}, \bibinfo {author}
  {\bibfnamefont {T.~J.}\ \bibnamefont {Lane}}, \bibinfo {author}
  {\bibfnamefont {G.}~\bibnamefont {Wetzstein}}, \bibinfo {author}
  {\bibfnamefont {P.}~\bibnamefont {Musumeci}}, \ and\ \bibinfo {author}
  {\bibfnamefont {D.}~\bibnamefont {Ratner}},\ }\bibfield  {title} {\enquote
  {\bibinfo {title} {Electron ghost imaging},}\ }\href {\doibase
  10.1103/PhysRevLett.121.114801} {\bibfield  {journal} {\bibinfo  {journal}
  {Phys. Rev. Lett.}\ }\textbf {\bibinfo {volume} {121}},\ \bibinfo {pages}
  {114801} (\bibinfo {year} {2018})}\BibitemShut {NoStop}%
\bibitem [{\citenamefont {Boltz}\ \emph {et~al.}(2018)\citenamefont {Boltz}
  \emph {et~al.}}]{Boltz18}%
  \BibitemOpen
  \bibfield  {author} {\bibinfo {author} {\bibfnamefont {T.}~\bibnamefont
  {Boltz}} \emph {et~al.},\ }\bibfield  {title} {\enquote {\bibinfo {title}
  {{Studies of longitudinal dynamics in the micro-bunching instability using
  machine learning}},}\ }in\ \href@noop {} {\emph {\bibinfo {booktitle} {{9th
  International Particle Accelerator Conference (IPAC 2018) Vancouver, BC
  Canada, April 29-May 4, 2018}}}}\ (\bibinfo {year} {2018})\ p.\ \bibinfo
  {pages} {THPAK030}\BibitemShut {NoStop}%
\bibitem [{\citenamefont {Nawaz}\ \emph {et~al.}(2018)\citenamefont {Nawaz},
  \citenamefont {Lichtenberg}, \citenamefont {Pfeiffer},\ and\ \citenamefont
  {Rostalski}}]{Nawaz:IPAC2018-WEPMF058}%
  \BibitemOpen
  \bibfield  {author} {\bibinfo {author} {\bibfnamefont {A.S.}\ \bibnamefont
  {Nawaz}}, \bibinfo {author} {\bibfnamefont {G.}~\bibnamefont {Lichtenberg}},
  \bibinfo {author} {\bibfnamefont {S.}~\bibnamefont {Pfeiffer}}, \ and\
  \bibinfo {author} {\bibfnamefont {P.}~\bibnamefont {Rostalski}},\ }\bibfield
  {title} {\enquote {\bibinfo {title} {{Anomaly Detection for Cavity Signals -
  Results from the European XFEL}},}\ }in\ \href@noop {} {\emph {\bibinfo
  {booktitle} {{9th International Particle Accelerator Conference (IPAC 2018)
  Vancouver, BC Canada, April 29-May 4, 2018}}}}\ (\bibinfo {year}
  {2018})\BibitemShut {NoStop}%
\bibitem [{Note1()}]{Note1}%
  \BibitemOpen
  \bibinfo {note} {\protect \url
  {https://www.nvidia.com/en-us/data-center/tesla/}}\BibitemShut {NoStop}%
\bibitem [{Note2()}]{Note2}%
  \BibitemOpen
  \bibinfo {note} {\protect \url
  {https://colab.research.google.com}}\BibitemShut {NoStop}%
\bibitem [{Note3()}]{Note3}%
  \BibitemOpen
  \bibinfo {note} {\protect \url
  {https://aws.amazon.com/machine-learning/}}\BibitemShut {NoStop}%
\bibitem [{Note4()}]{Note4}%
  \BibitemOpen
  \bibinfo {note} {\protect \url
  {https://cloud.google.com/ml-engine/}}\BibitemShut {NoStop}%
\end{thebibliography}%

\end{document}